\documentclass[12pt,notoc,hyper]{JHEP3}

\usepackage{graphicx}
\usepackage{dcolumn}
\usepackage{bm}

\usepackage{epsfig,multicol,bbm} 

\def\simlt{\lower.5ex\hbox{$\; \buildrel < \over \sim \;$}}
\def\simgt{\lower.5ex\hbox{$\; \buildrel > \over \sim \;$}}

\title{Single-field inflation constraints from CMB and SDSS data}

\author{Fabio Finelli\\
INAF/IASF Bologna, Istituto di Astrofisica Spaziale e Fisica Cosmica di
Bologna, via Gobetti 101, I-40129 Bologna, Italy\\
INFN, Sezione di Bologna, via Irnerio 46, I-40126 Bologna, Italy\\
Email: \email{finelli@iasfbo.inaf.it}}
\author{Jan Hamann\\
Department of Physics and Astronomy, University of Aarhus, Ny Munkegade,
DK-8000 Aarhus C, Denmark\\
Email: \email{hamann@phys.au.dk}}
\author{Samuel M. Leach\\
SISSA, Astrophysics Sector, via Beirut 2-4, I-34151 Trieste, Italy\\
INFN, Sezione di Trieste, I-34151 Trieste, Italy\\
Email: \email{leach@sissa.it}}
\author{Julien Lesgourgues\\
CERN, Theory Division, CH-1211 Geneva 23, Switzerland\\
Institut de Th\'eorie des Ph\'enom\`enes Physiques, EPFL, CH-1015 Lausanne,
Switzerland\\
LAPTh (CNRS - Universit\'e de Savoie), BP 110, F-74941 Annecy-le-Vieux
Cedex, France\\
Email:\email{ julien.lesgourgues@cern.ch}}

\preprint{\astroph{0912.0522}, CERN-PH-TH/2009-241, LAPTH-1368/09}      

\abstract{
  We present constraints on canonical single-field inflation derived
  from WMAP five year, ACBAR, QUAD, BICEP data combined with the halo
  power spectrum from SDSS LRG7. Models with a non-scale-invariant
  spectrum and a red tilt $n_{\rm S}<1$ are now preferred over the
  Harrison-Zel'dovich model ($n_{\rm S}=1$, tensor-to-scalar ratio
  $r=0$) at high significance.  Assuming no running of the spectral
  indices, we derive constraints on the parameters ($n_{\rm S}$, $r$)
  and compare our results with the predictions of simple inflationary
  models. The marginalised credible intervals read $n_{\rm S} =
  0.962^{+0.028}_{-0.026}$ and $r<0.17$ ($95\%$ confidence level).
  With respect to previous analyses, the portion of the $68\%$~c.l.
  contours compatible with potentials which are concave in the
  observable region becomes even smaller, but the quadratic potential
  model remains inside the $95\%$~c.l. contours. We demonstrate that
  these results are robust to changes in the datasets considered and
  in the theoretical assumptions made.  We then consider a
  non-vanishing running of the spectral indices by employing different
  methods, non-parametric but approximate, or parametric but exact.
  With our combination of CMB and LSS data, running models are
  preferred over power-law models only by a $\Delta \chi^2 \simeq
  5.8$, allowing inflationary stages producing a sizable negative
  running $-0.063^{+0.061}_{-0.049}$ and larger tensor-scalar ratio
  $r<0.33$ at the $95\%$~c.l. This requires large values of the third
  derivative of the inflaton potential within the observable range. We
  derive bounds on this derivative under the assumption that the
  inflaton potential can be approximated as a third order polynomial
  within the observable range.  }


\begin{document}

\section{Introduction}

Recent advances in centimetre and millimetre wave detector technology
are ushering in an exciting observational window on the physics of the
early universe. The celebrated `inflation' model 
\cite{Starobinsky:1980te,Guth:1980zm}
provides a compelling
framework for thinking about conditions that might have prevailed in
the early universe, yet to date there is no real consensus on what is
the underlying inflationary model. A well-known implementation of
the idea relies on scalar field models, for which the
inflationary predictions can, nonetheless, be worked out in detail.
Although it is sometimes claimed that inflation `can predict
anything', this view misses the point that, notwithstanding the
impressive successes of the Wilkinson Microwave Anisotropy Probe
(WMAP) \cite{2009ApJS..180..225H}, currently there is a real dearth
of hard new facts relating to early universe physics, and that the
predictions of inflation have yet to be put under any real strain.

As a consequence there is a need for continual monitoring of the
parameters of a generic inflation model, as constrained primarily by
measurements of Cosmic Microwave Background (CMB) anisotropies and
surveys of the tracers of Large Scale Structure (LSS). Once the values
of the parameters of inflation have been determined then the
implications for the physics of inflation can be worked out. In the
long term, it is possible that an improved understanding of the
physics of inflation could put an important constraint on the physics
of the early universe, just as an understanding of the physics of
nucleosynthesis has for a long time been an important constraint on
physics beyond the standard model of particle physics.

Some constraints on canonical inflation\footnote{For recent observational
  constraints on models with non-canonical kinetic terms, see e.g.
  \cite{Lorenz:2008je,Agarwal:2008ah,Tzirakis:2008qy}. For
  constraints on canonical models incorporating WMAP5 data, one should
  add to the list above the update of Ref.~\cite{2006JCAP...08..009M} 
  published in the book \cite{Schwarz:2008zz}.}
incorporating WMAP five-year data (WMAP5) have
been published by Komatsu et al. \cite{Komatsu:2008hk}, Peiris \&
Easther \cite{Peiris:2008be} and Kinney et al.
\cite{Kinney:2008wy}, using different combinations of WMAP,
small-scale CMB data from ACBAR~\cite{2009ApJ...694.1200R}, Baryon
acoustic oscillation (BAO) data from~\cite{Percival:2007yw}, or SNIa
data from the Supernovae Legacy Survey~\cite{Astier:2005qq}.  Since
then, some important new data sets have been released, like CMB
polarisation measurements from BICEP~\cite{2009arXiv0906.1181C} and
QUaD~\cite{2009arXiv0906.1003Q}, the galactic halo correlation
function derived from Luminous Red Galaxies in the data release seven
of the of the Sloan Digital Sky Survey (LRG7), and the luminosity
distance of Type Ia supernovae (SNIa) from the SDSS-II survey
\cite{Kessler:2009ys}. These new data sets have been a stimulus to
our analysis at a time when increasingly detailed measurements of the
CMB from WMAP and Planck~\cite{2006astro.ph..4069T}, as well as a number of
sub-orbital experiments such as ACT~\cite{2009arXiv0907.0461H},
SPT~\cite{2004SPIE.5498...11R}, QUIET~\cite{2008arXiv0806.4334S},
SPIDER~\cite{2008SPIE.7010E..79C},
PolarBear~\cite{2008AIPC.1040...66L} and
EBEX~\cite{2004SPIE.5543..320O}, are expected over the course of the
next couple of years.

In this work, we will use a combination of the most up-to-date CMB
data and of the LRG7 halo correlation function 
for updating constraints on single-field inflationary models. After
discussing our methodology in Section~\ref{methodology}, we present a
discussion of the status of the Harrison-Zel'dovich model in
Section~\ref{HZ}, which is the simplest possible empirical model for
primordial perturbations. Then, in Section~\ref{simple}, we show our
results for the simplest class of inflationary models, those that do
not lead to any significant running of the tilt in the scalar
primordial spectrum. In Section~\ref{conservative}, we take a more
conservative point of view of deriving constraints on the observable
part of the inflaton potential using different methods (non-parametric
but approximate, or parametric but exact). Our conclusions are
summarized in Section~\ref{conclusions}.

\section{Methodology\label{methodology}}

We perform an analysis of the WMAP
\cite{2009ApJS..180..296N,2009ApJS..180..306D},
ACBAR~\cite{2009ApJ...694.1200R}, BICEP \cite{2009arXiv0906.1181C} and
QUaD \cite{2009arXiv0906.1003Q} CMB anisotropy data, combined with the
LRG7 halo correlation function computed in
\cite{2009arXiv0907.1659R,2009ApJ...702..249R}. Our precise choice of
the following subsets of data, hard won by all of these experiments,
reflects our judgments and decisions that were necessary in order to
avoid double counting of data.  

Firstly, we prefer to use cosmic variance limited (CVL) data from a
satellite experiment, where available from WMAP which is CVL up to
$\ell=530$~\cite{2009ApJS..180..296N}, which can be judged to be data
of the very highest quality. Note that all suborbital experiments
derive their absolute calibration from cross-calibration off the WMAP
temperature anisotropies and ultimately from the CMB dipole, which
more broadly speaking can be taken as an indication that the
suborbital experimental teams agree with our assessment of the quality
of WMAP data. For this reason we remove a) the ACBAR band powers with
$\ell<790$ and $\ell>1950$ to avoid overlap with WMAP and to avoid
contamination from potential foreground residuals, and b) the QUAD TE
bandpowers 1--4, and the BICEP TT and TE bandpowers which again
overlap excessively with WMAP.

A slightly less appreciated second point is the overlap of QUAD and
ACBAR. QUAD was originally optimized for making CMB polarization
measurements~\cite{2004MNRAS.349..321B}, and so they chose to observe
a small region of sky, and one that overlapped with the `Boomerang
deep field' in order to improve their absolute calibration inherited
from WMAP.  ACBAR also observed the same field, which they called
their `CMB region 8' (see Figure 1 and Table 1
of~\cite{2009ApJ...694.1200R}). In terms of gaining independent
constraints on the CMB temperature spectrum, it is rather unfortunate,
but perhaps unavoidable, that both ACBAR and QUAD observed the same
region of sky. In our final analysis, we chose to drop the QUAD
temperature data because ACBAR have greater statistics on the CMB
temperature spectrum (as is evident from their smaller power spectrum
errors bars) having observed a wider area of sky in total, better
optimized for making temperature (rather than polarization)
measurements.

Unless otherwise specified, all the results of this paper are derived
from this combination of data.  In some particular cases, we also
considered the impact of ancillary data such as the luminosity
distance of SNIa from \cite{Kessler:2009ys} and the recent
determination of the Hubble parameter \mbox{$H_0=74.2\pm3.6$ km
s$^{-1}$ Mpc$^{-1}$} by \cite{2009ApJ...699..539R}.

We use \texttt{CosmoMC} \cite{2002PhRvD..66j3511L,2000ApJ...538..473L}
in order to compute the Bayesian probability distribution of model
parameters. The pivot scale of the primordial scalar and tensor power
spectra was set to $k_*=0.017$~Mpc$^{-1}$, as recommended by
\cite{2007PhRvD..75h3520C}, and we have verified that this choice of
pivot scale is still close to optimum for the combination of data used
in our analysis (see Section~\ref{conservative}).  Apart from the
primordial spectrum parameters (or inflationary parameters) described
in the next section, we vary the baryon density $\omega_{\rm
b}=\Omega_{\rm b} h^2$, the cold dark matter density $\omega_{\rm c}=
\Omega_{\rm c}h^2$, the Hubble parameter $H_0=100 h \, {\rm km}\,{\rm
s}^{-1}{\rm Mpc}^{-1}$ and the reionisation optical depth $\tau$. We
assume a flat universe, and so the cosmological constant for each
model is given by the combination $\Omega_{\Lambda} = [1-(\omega_{\rm
b}+\omega_{\rm c})h^{-2}]$. We also assume a CMB temperature $T_{\rm
CMB}=2.725$~K~\cite{2009arXiv0911.1955F} and three neutrinos with a
negligible mass (excepted in subsection \ref{neutrinomass}).  We set
the primordial helium fraction to $y_{\rm He}=0.248$, a value
consistent with the predictions of standard Big Bang Nucleosynthesis
for a baryon density of $\omega_{\rm b} \simeq 0.0022$.\footnote{Under
the assumptions of standard BBN, the commonly used value of $y_{\rm
He}=0.24$ corresponds to $\omega_{\rm b} \sim 0.01$, which is clearly
inconsistent with observations.  With the rather modest sensitivity of
present data to $y_{\rm He}$ \cite{2008PhRvD..78d3509I}, this choice
may not lead to significant bias, but for data from the Planck
satellite the issue will become relevant~\cite{2008JCAP...03..004H}.}
In order to fit WMAP, ACBAR and QUaD data, we use the lensed CMB and
matter power spectra and we follow the method implemented in
\texttt{CosmoMC} consisting in varying a nuisance parameter $A_{\rm
SZ}$ which accounts for the unknown amplitude of the thermal SZ
contribution to the small-scale CMB data points assuming the model of
\cite{2002MNRAS.336.1256K}.  We use a \texttt{CAMB} accuracy setting
of at least $1.2$.  We sample the posterior using the
Metropolis-Hastings algorithm \cite{Hastings:1970xy} at a temperature
$T=2$ (for improved exploration of the tails), generating eight
parallel chains and imposing a conservative Gelman-Rubin convergence
criterion \cite{GelmanRubin} of $R-1 < 0.01$.

\section{Discussion of HZ model\label{HZ}}

The empirical model for the spectrum of primordial perturbations
proposed by Harrison \cite{1970PhRvD...1.2726H}, Zel'dovich
\cite{1972MNRAS.160P...1Z} and Peebles \cite{1970ApJ...162..815P}
(hereafter the HZ spectrum) has a single parameter describing
primordial perturbations, namely the primordial spectrum amplitude
$A_{\rm S}$. It is interesting to compare its goodness-of-fit with the
next-level model in terms of number of free parameters, namely a model
with a power-law primordial spectrum with tilt $n_{\rm S}$ but no tensor
perturbations (this model can be motivated by low-scale slow-roll
inflation).
The general mood is that with the increasing precision of CMB
anisotropy measurements, the HZ spectrum is on the verge of being
ruled out in favour of a model with a red tilt.

However, keeping an open mind about the origin of primordial
perturbations, the HZ spectrum still holds an allure as a possible
indicator of to-be-discovered symmetries of the early universe. As a
result we have taken a slightly different tack by assuming for a
moment that the HZ spectrum is not only a viable but a correct
description of the data, thereby reducing the dimensionality of the
parameter space, in order to examine in detail some of the
cracks in this model that are beginning to appear.
Using our basic data set (WMAP5, ACBAR, BICEP, QUaD and LRG7) we ran
MCMC chains for the HZ model and for the tilted model.  Introducing a
tilt decreases the minimum effective chi square by $\Delta (- 2
\ln({\cal L})) = 12.6$, which clearly implies a tension between the HZ
model and the data\footnote{While this work was being completed,
Refs.~\cite{2009arXiv0911.5191K,Peiris:2009wp} appeared and adressed
this issue more specifically with different methods.  In
Ref.~\cite{2009arXiv0911.5191K}, the Bayesian evidence ratio between
the HZ and power-law models is computed for a data set slightly
different from ours, using a prior $0.8 < n_{\rm S} < 1.2$ for the
power-law model. The authors conclude that there is strong negative
evidence for the HZ model. In this paper we do not quote an evidence
ratio because this would require a very good sampling of the distribu-
tion tail for ${\cal P}(n_{\rm S})$ in the power law model, four
sigma away from the maximum, and our runs were not designed for this
purpose. Instead, Ref.~\cite{Peiris:2009wp} proposes a method for
reconstructing the primordial scalar spectrum with no underlying
theoretical or inflationary prior. The conclusion of this work is
still that a scale-invariant spectrum is disfavored, but only weakly
(consistently, when thoretical priors are removed, one reaches more
conservative conclusions).}. On top of that, the HZ model also leads
to different preferred ranges for the cosmological parameters, as can
be checked from Figure~\ref{fig:hz} and Table \ref{tab:hz}. In
particular, fixing $n_{\rm S}=1$ implies a rather high baryon
density of $\omega_{\rm b } =0.0239 \pm 0.0007$ (at 95\%~c.l.), which
is at odds with recent constraints from measurements of primordial
element abundances within the standard Big Bang Nucleosynthesis
scenario. Indeed, a conservative bound based only on the deuterium
abundance yields $\omega_{\rm b}=0.0213\pm 0.002$ (at 95\%~c.l.)
\cite{2008MNRAS.391.1499P}. The statistical error shrinks by a factor
two when $^4$He measurements are included
\cite{2009PhR...472....1I}. Taking the deuterium bound on
$\omega_{\rm b}$ as a prior yields $\Delta (- 2 \ln({\cal L})) =
17.6$ -- a difference exceeding the 4$\sigma$ level. We conclude that
the HZ model is now clearly incompatible with CMB and LSS data under
the assumption of the standard $\Lambda$CDM cosmology.

We remark that the HZ model also implies a value of the baryon
fraction $f_{\rm b} \equiv {\omega_{\rm b}/(\omega_{\rm b} +
\omega_{\rm c})}= 0.173 \pm 0.005$ (at $68\%$ c.l.) which is about a
percent higher than the corresponding value in the $n_{\rm S}$ model
which impliess $f_{\rm b} = 0.162 \pm 0.005$ (at $68\%$ c.l.). It
appears though that, notwithstanding observational uncertainties of
the total mass of clusters and theoretical uncertainties in the
physics of feedback in clusters, the slightly higher value of the
baryon fraction in the HZ model can easily be accommodated by the most
recent X-ray observations of clusters (see for instance
\cite{2009A&A...498..361P}). This result merely highlights the
current, but rather mild, model dependence of the baryon fraction.

\FIGURE{
\centering
\includegraphics[height=.75\textwidth,angle=270]{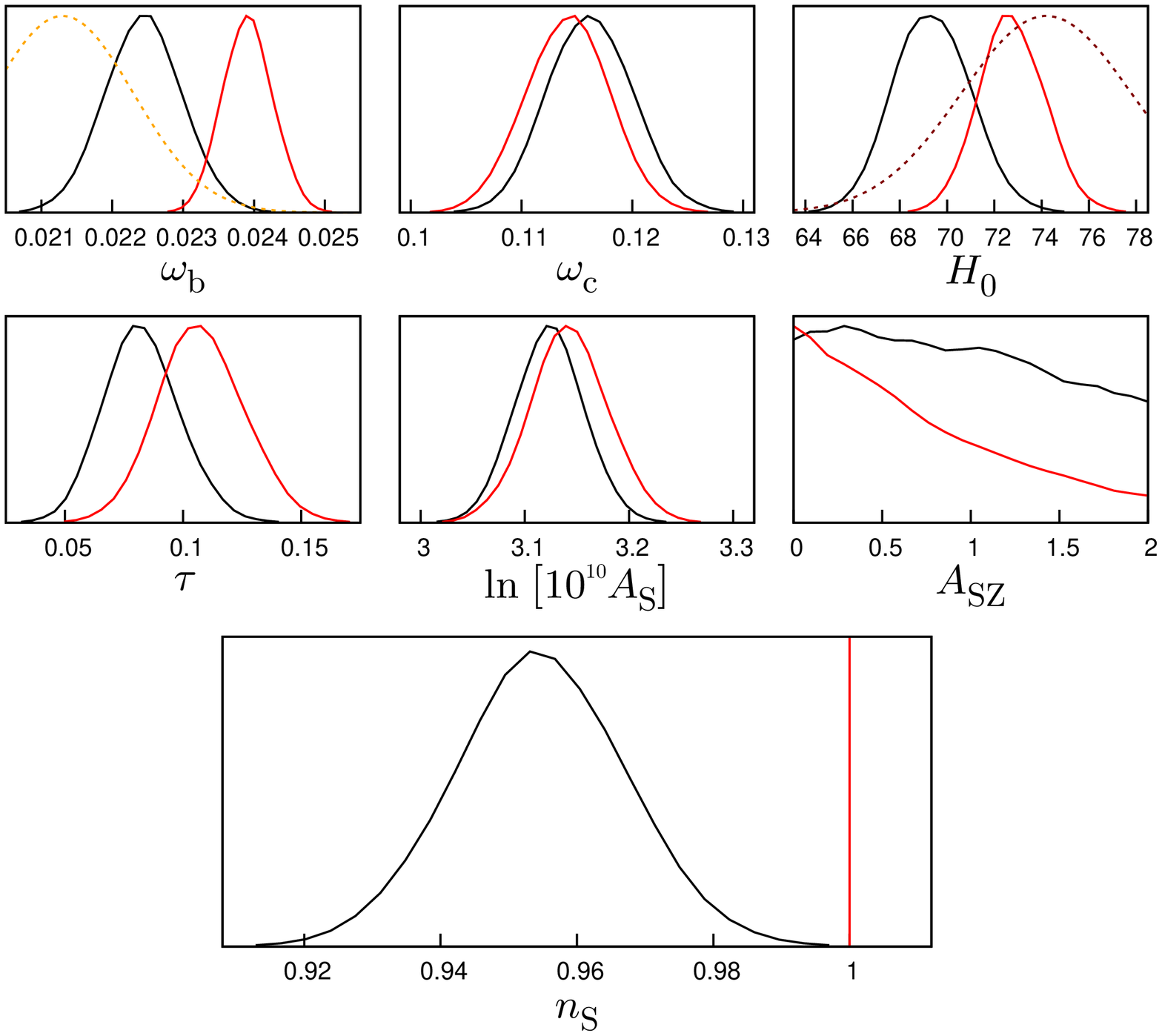}
\caption{Posterior probabilities of the parameters of the HZ model
  (red) and the tilted model (black). The dotted orange line denotes
  constraints on $\omega_{\rm b}$ from BBN and the dotted maroon line
  represents the HST prior on $H_0$.}
\label{fig:hz}
}

\TABLE{
\centering
\begin{tabular}{|l|cc|}
\hline
& HZ & tilted \\
\hline
$\omega_{\rm b}$ & $0.0239 \pm 0.0007$ & $ 0.0224 \pm 0.001$ \\
$\omega_{\rm c}$ & $0.114 \pm 0.007$    &   $0.116 \pm 0.007  $  \\
$h$ &  $0.728 ^{+0.027}_{-0.026} $ &  $0.694 ^{+0.032}_{-0.030} $   \\
$\tau$ & $0.108 ^{+0.036}_{-0.034}$  &   $0.083 ^{+0.032}_{-0.029}$ \\
$\log \left[ 10^{10} A_{\rm S}\right]$ & $3.14 \pm 0.07 $  & $3.12 \pm 0.06 $ \\
$n_{\rm S}$ &--& $0.955^{+0.024}_{-0.026} $  \\
\hline
$\sigma_8$ & $0.853 \pm 0.048$  & $0.822^{+0.047}_{-0.045} $\\
\hline
\end{tabular}
\caption{\label{tab:hz}
  Mean parameter values and bounds of the central 95\%-credible intervals
  for the HZ model and the tilted model.  }
}
Finally we note from Figure~\ref{fig:hz} that the
preferred value of $A_{\rm SZ}$ is zero, which results from the fact
that both $n_{\rm S}$ and $A_{\rm SZ}$ change the ratio of large-scale
to small-scale power in the $C_{\ell}$. Taken at face value this is
arguably another point against the HZ model. On the other hand the HZ
model does prefer a value of $H_0$ that is slightly more consistent
with value of $H_0 = 74.2 \pm 3.6$ km s$^{-1}$ Mpc$^{-1}$
from~\cite{2009ApJ...699..539R}.

To summarise this section, we have argued that the HZ
  model faces pressure both on grounds of goodness of fit, but also on
  grounds of poor astrophysical consistency. As
  mentioned later in section \ref{neutrinomass}, the inclusion of 
  neutrino masses would not alleviate the pressure on the HZ model.

\section{Constraints on the simplest models of inflation\label{simple}}

\subsection{Motivations\label{motivnorunning}}

Most inflationary model discussed in the literature lead to a
power-law scalar spectrum, i.e., to a negligible running of the scalar
tilt. This follows from a well-known argument which can be summarised
in the following way. The function accounting for the Hubble parameter
as a function of the field (or as a function of the number of
$e$-folds) can be expanded around any value in an infinite hierarchy
of slow-roll parameters $\epsilon_n$, each one accounting for the
logarithmic derivative of the previous term $\epsilon_{n-1}$ (see for
instance \cite{Schwarz:2001vv,Leach:2002ar,Kinney:2002qn} or Appendix
\ref{A2}). Current bounds on the tensor-to-scalar ratio $r$ and
scalar tilt $n_{\rm S}$ imply that the first two parameters
$\epsilon_1$ and $\epsilon_2$ are much smaller than one, as expected
during slow-roll inflation.  So, a sizeable running $\alpha_{\rm S}$
on observable scales can only be generated when one of the parameters
$\epsilon_n$ with $n \geq 3$ is roughly of order one or larger when
observable scales leave the Hubble radius during inflation, i.e., when
some derivatives of the first two slow-roll parameters are very large.
If this is not the case, the running will be of the order of
$\epsilon_1 \epsilon_2$, i.e., at most of the order of $\alpha_{\rm S}
\sim {\cal O}(10^{-4})$.

On the other hand, the number of $e$-folds required between that time
and the end of inflation (which depends on the scale of inflation, on
the details of inflation ending, and on the efficiency of the
reheating mechanism) can be conservatively assumed to lie in the range
from 30 to 60 $e$-folds. For any smooth potential, a sizeable running
$|\alpha_{\rm S}| > 0.01$ implies such large derivatives of the first
two slow-roll parameters in the observable range that inflation would
end very few $e$-folds after galaxy scales leave the Hubble radius,
not even reaching 30 inflationary $e$-folds
\cite{Makarov:2005uh,Easther:2006tv}. This argument can only be
evaded: (i) with very special potentials (incoporating sharp features,
or such that their Taylor-expansion involves very high-order
coefficients, see for instance
\cite{Makarov:2005uh,Ballesteros:2005eg,Ballesteros:2007te}); (ii)
within set-ups involving several inflaton fields, leading to a phase
transition or even to several short stages of inflation, such that the
required 30 to 60 inflationary $e$-folds are not contiguous. Since
these situations are beyond minimal requirements, and since the
sensitivity of current data to running ($\Delta \alpha_{\rm S} \sim
{\cal O}(10^{-2})$) is far too small for probing
generic slow-roll model predictions
($\alpha_{\rm S} \leq {\cal O}(10^{-4})$), it appears sensible to
perform an analysis where the running is assumed to be negligible, in
order to preserve the simplicity of the inflationary paradigm. We will
relax this theoretical prejudice in the next section.

We thus consider all inflationary models which can be described by the
primordial perturbation parameters consisting of the scalar amplitude and
spectral index $(A_{\rm S}, n_{\rm S})$, and the tensor-to-scalar
ratio $r$ (both defined at the pivot scale $k_*=0.017$~Mpc$^{-1}$).
In single-field inflation, deep in the slow-roll limit, the tensor
spectrum shape is not independent of the scalar one. We will consider
a tensor spectrum with a tilt $n_{\rm T}=-r/8$, as predicted for
canonical single-field inflation at first-order in slow-roll.

\subsection{Results for the basic data set, and implications for inflation}

In Table \ref{tab:vanilla+r}, we present constraints on each parameter
of this model, using only CMB data in the first column, and our
reference data set (CMB plus LRG7) in the second one. In Figure
\ref{fig:IVA} we focus on the joint probability in the $(n_{\rm S},
r)$ plane, for the reference data set.  Compared to the results
obtained by \cite{Komatsu:2008hk}, our constraints are shifted towards
slightly lower tilts and significantly lower tensor-to-scalar ratios.
The marginalised 95\%-credible interval for the tilt is given by
$n_{\rm S}=0.962^{+0.028}_{-0.026}$ (to be compared with the result
of the previous section, $n_{\rm S}=0.955^{+0.024}_{-0.026}$,
obtained for low energy scale inflationary models with $r \simeq 0$).
At the same confidence level, our result for the tensor-to-scalar
ratio is $r < 0.17$. This corresponds to an upper bound on the energy
scale of inflation
\begin{equation}
V_* = \frac{3 A_{\rm S} \, r \, m_{\rm P}^4}{128} < 8.9 \times 10^{-12}
m_{\rm P}^4
= (2.1 \times 10^{16} {\rm GeV} )^4 \nonumber
\end{equation}
at 95\%~c.l., where $m_{\rm P}=G^{-1/2}$ stands for the Planck mass
(the reduced Planck mass, referred to in the next
  sections, will be denoted as $M_{\rm P}=(8 \pi G)^{-1/2}$). Bounds on quantities which are not sharply constrained
  by the data, like $r$ or $V_0$, depend mildly on the choice of
  parametrisation, as illustrated in \cite{Valkenburg:2008cz}, but for
  reasonable choices this will not affect our qualitative
  conclusions.

\TABLE{
\centering
\begin{tabular}{|l|ccc|}
\hline
& CMB only & CMB+LRG (DR7) & CMB+BAO (DR7)  \\
\hline
$\omega_{\rm b}$ & $0.0230 ^{+0.0013}_{-0.0012}$ & $ 0.0227 \pm 0.0011$ &  $0.0226 \pm 0.0011$ \\
$\omega_{\rm c}$ & $0.105  ^{+0.012}_{-0.013}$    &   $0.115 ^{+0.008}_{-0.007}$ & $0.113 \pm 0.007$ \\
$h$ &  $0.746 ^{+0.066}_{-0.056} $ &  $0.700 ^{+0.034}_{-0.032}$  & $0.708 \pm 0.030$ \\
$\tau$ & $0.091 ^{+0.036}_{-0.034}$  &   $0.083 ^{+0.033}_{-0.029}$ & $0.085^{+0.033}_{-0.030} $ \\
$\log \left[ 10^{10} A_{\rm S}\right]$ & $3.08 \pm 0.08 $  & $3.12^{+0.06}_{-0.07} $ & $3.11^{+0.07}_{-0.06} $ \\
$n_{\rm S}$ & $0.975  ^{+0.039}_{-0.031} $ & $0.962^{+0.028}_{-0.026}$ & $0.961^{+0.027}_{-0.026}$ \\
$r$ & $< 0.29$ & $< 0.17$ & $< 0.18$ \\
\hline

\end{tabular}
\caption{\label{tab:vanilla+r}
  Mean parameter values and bounds of the central 95\%-credible intervals
  for the parameters of the vanilla$+r$ model and various combinations
  of data sets.  For the tensor-to-scalar ratio $r$ the 95\%-credible
  upper bound is quoted.}
}

\FIGURE{
\centering
\includegraphics[height=.66\textwidth,angle=270]{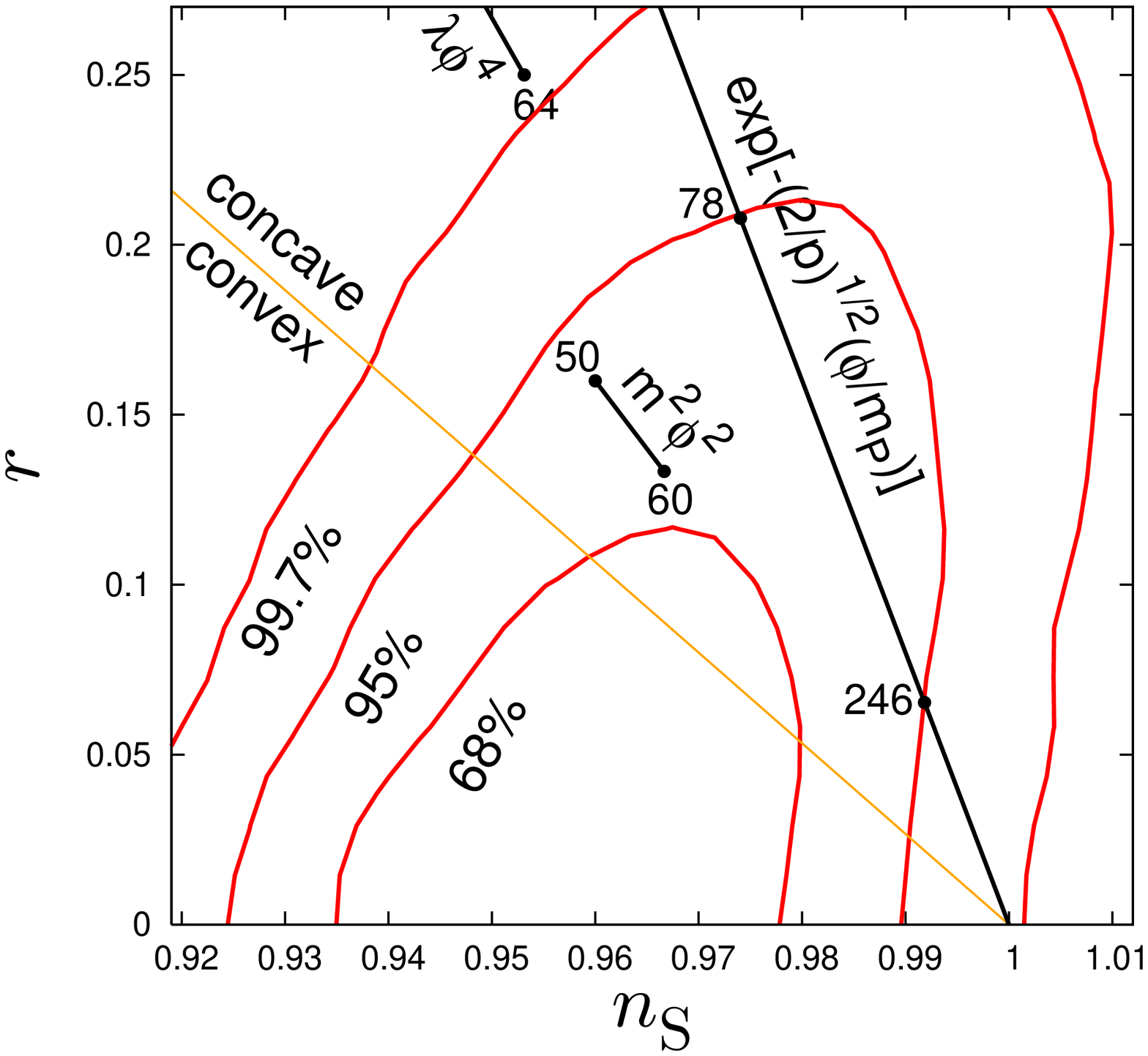}
\caption{\label{fig:IVA} Marginalised joint 68\%-, 95\%- and
  99.7\%-credible contours in the ($n_{\rm S}$, $r$) plane for our
  reference data set (CMB plus LRG7). The orange line shows the limit
  between inflationary models with a concave/convex potential in the
  observable range. We show in black the regions corresponding to
  popular concave models with a quadratic, quartic or exponential
  potential. }
}

As pointed out in previous works (see for instance
\cite{Komatsu:2008hk,Kinney:2008wy}), current data mainly raise some
tension for inflationary models with a concave potential in the
observable range. Most of the 68\% allowed region lies below the
convex potential limit, and convex models with a red tilt in the range
[0.93-0.99] are comfortably allowed by the data. These include for
instance:
\begin{itemize}
\item `Natural inflation'~\cite{1990PhRvL..65.3233F} with the potential $V(\phi) = \Lambda^4
  \left[ 1+\cos(\phi/f)\right]$ leading to the prediction $n_{\rm S}
  \simeq 1 - m^2_{\rm P}/(8 \pi f^2)$, which is allowed only for $f > 0.73
  \, m_{\rm P}$ (95\%~c.l.).
\item
More generally, `new inflation' type models where the inflaton rolls
away from unstable equilibrium with a slope given by $V(\phi) = V_0 [1
- (\phi/\mu)^{\alpha}]$. For $\alpha=2$ and assuming $|\phi| \ll \mu$
when observable scales leave the Hubble radius, our constraints imply
$1.5 < \mu/m_{\rm P} < 4.0$ (95\%~c.l.). 
Inflationary models with $\alpha \ge 3$ are all in good agreement with
observations. 
\item
The symmetry-breaking  
potential $V(\phi) = \lambda (\phi^2 -\phi_0^2)^2$ 
\cite{Olive:1989nu,Boyanovsky:2009xh} 
is convex in its observable region for
$\phi_0<6.6 \, m_{\rm P}$. We find that this model
is in agreement with observations for 
$\phi_0>2.1 \, m_{\rm P}$ and  $\lambda \sim {\cal O} (10^{-14})$.
\item `Hybrid inflation' with a logarithmic slope (caused for instance
  by one-loop corrections during global SUSY inflation
  \cite{Dvali:1994ms}), which predicts a small $r$ and $n_{\rm S}$
  close to 0.98.

\item
Inflationary models predicting a red tilt for scalar perturbations and
a tensor-to-scalar ratio $\sim {\cal O} ((n_{\rm S}-1)^2)$, as $R +
R^2/(6 M^2)$ \cite{Starobinsky:1980te} or Standard Model Higgs
inflation \cite{Bezrukov:2009db}, are in good agreement with
observations
(as well as models with an even lower $r$, see e.g.
Refs.~\cite{Allahverdi:2006we,Ross:2009hg}).

\end{itemize}
Among concave models, we can single out a few generic models for
further discussion:
\begin{itemize} 

\item Inflation with a quadratic potential $V=\frac{1}{2}m^2 \phi^2$
  (often dubbed `chaotic inflation' after Ref.~\cite{Linde:1983gd})
  lives along the line $r \simeq -4 (n_{\rm S}-1)$ during slow-roll.  The
  position of the model along this line depends on the number of
  $e$-folds $N_*$ between the time of Hubble crossing for the pivot
  scale, and the end of inflation. In turn, this number depends on the
  reheating temperature. Due to uncertainties on the reheating stage,
  we consider here a plausible range $50<N_*<60$, for which quadratic
  inflation is still within the 95\%-credible region in the
  ($n_{\rm S}$, $r$) plane, as can be seen in Figure~\ref{fig:IVA}.
  Note that updated constraints on the mass can be inferred from the
  combination $(3 \pi A_{\rm S})^{1/2} / (2 N_*)$ which is equal to
  $(m/m_{\rm P})$ in this model. Varying $A_{\rm S}$ in the
  95\%-credible range given by Table \ref{tab:vanilla+r}, and $N_*$
  between 50 and 60, we find that the mass of chaotic inflation models
  lies in the range from $1.2 \times 10^{-6} m_{\rm P}$ to $1.5
  \times 10^{-6} m_{\rm P}$.

\item Inflation with a quartic potential $V=\lambda \phi^4/4$ lives
  along the line $r = -16/3 (n_{\rm S}-1)$ during slow-roll.  For this
  model, there are indications that the number of $e$-folds $N_*$ should
  be increased by four, for the reasons summarised in
  \cite{Liddle:2003as}.  In Figure~\ref{fig:IVA} we show the line
  corresponding to quartic inflation for $N_*<64$. This model is found
  to be outside of the 99.7\%-credible region in the ($n_{\rm S}$,
  $r$) plane.

\item Inflation with an exponential potential $V=\exp[ -
  \sqrt{p/2}(\phi/M_P)]$ is called power-law inflation \cite{1985PhRvD..32.1316L}, 
  because the
  exact solution for the scale factor is given by $a(t) \propto t^p$.
  This model is incomplete, since inflation would not end without an additional mechanism 
  which stops it. Assuming that such a
  mechanism exists and leaves unmodified its predictions on
  cosmological perturbations, we can constrain this model since it
  predicts $r=-8 (n_{\rm S}-1)$ and $n_{\rm S} = 1 - 2/(p-1)$. Only models with
  $78 < p < 246$ are found to lie within our 95\%-credible region in
  the ($n_{\rm S}$, $r$) plane.

\item Hybrid or `false-vacuum' inflation with a quadratic slope and an
  effective potential $V= V_0 + \frac{1}{2}m^2 \phi^2$ is also
  compatible with the data provided that $V_0$ is not too large with
  respect to the quadratic term.  In this model, the two
    slow-roll parameters are related through $\alpha \eta^2 + \epsilon
    -\eta =0$, where $\alpha \equiv 8 \pi V_0 / (m^2
    m_\mathrm{P}^2)$. For $\alpha > 80$ this relation is not
    compatible with our 95\%-credible region in $(n_\mathrm{S}, r)$
      space. We conclude that independently of the mechanism
      responsible for the end of hybrid inflation, this class of
      models should obey to the constraint $V_0 \leq 3 m^2
      m_\mathrm{P}^2$.  Note however that this model is usually
      invoked as a way to obtain $|\phi_*| \ll M_{\rm P}$, and to avoid
      large radiative corrections during inflation
      \cite{Lyth:1998xn}. This can be achieved when the term $V_0$
      dominates over the quadratic one, i.e., in the large $\alpha$
      limit -- a situation constrained by current data.

\end{itemize}

\subsection{Impact of extra data sets}

\subsubsection{Baryon acoustic oscillations}
The results presented in the third column of Table \ref{tab:vanilla+r}
show that for the model considered in this section,
the constraints on the primordial spectrum parameters do not
degrade significantly if we replace the LRG7 power spectrum by
constraints on the baryon acoustic oscillation scale derived
from the same survey data~\cite{2009arXiv0907.1660P}.  This is a clear
indication that the shape information contained in the galaxy power
spectrum data is not very relevant here -- the improvement over the
results from a CMB only analysis stems entirely from the geometrical
information; it also implies that our estimates are unlikely to be
affected by uncertainties in the modeling of non-linear structure
growth at small scales.

\subsubsection{Supernova luminosity distances}

Even though the luminosity distances of type Ia supernovae are not a
direct probe of the primordial perturbations, their ability to
constrain the expansion history of the low-redshift universe may still
contribute useful information about inflation by helping alleviate
parameter degeneracies in the model.  The most recent compilation of
SN data was presented by Kessler et al.~\cite{Kessler:2009ys}, who
give two different sets of luminosity distances, derived from the same
observations, but using either the {\sc mlcs2k2} or {\sc salt-ii}
lightcurve-fitting algorithms.  In order to test whether SN add any
useful information to our analysis of inflationary parameters we have
separately combined the two luminosity distance data sets with the
CMB+LRG7 data.  In Figure~\ref{fig:nsr} we plot the resulting
constraints in the ($n_{\rm S}$, $r$) plane.  The discrepancy between
the results for the two light curve fitters is evident, and has
already been pointed out in Ref.~\cite{Kessler:2009ys} in the context
of the dark energy equation of state parameter. Whereas adding the
{\sc salt-ii} data has no appreciable effect on our $n_{\rm S}$ and
$r$ constraints, the {\sc mlcs2k2} data would lead to a significantly
tighter bound on the tensor-to-scalar ratio.  This effect can be
traced back to {\sc mlcs2k2}'s apparent preference for large values of
$\Omega_m$.  Given that at the moment the SN data fail a basic
self-consistency test between different methods, we choose to refrain
from combining them with the more robust CMB+LRG7 data in the
following sections.

\subsubsection{Direct constraints on the Hubble parameter}

We studied the impact of imposing the recent determination of $H_0$
from the measurement of nearby supernovae with the Hubble Space
Telescope~\cite{2009ApJ...699..539R}, which corresponds closely
to $H_0 = 74.2 \pm 3.6$ km s$^{-1}$ Mpc$^{-1}$ for the $\Lambda$CDM
cosmology (we verified that `post-processing' the Markov chains with a
Gaussian prior on $H_0$ yields virtually identical results to those
obtained using the weakly cosmology dependent $H_0$ likelihood code
provided by \cite{Percival:2007yw} and included in the October 2009
version of \texttt{CosmoMC}).  We found that since CMB+LRG7 prefer a
slightly lower value of $H_0$ at 1$\sigma$, then imposing the $H_0$
constraint shifts $\omega_{\rm B}$ and hence $n_{\rm S}$ to slightly
larger values.

\subsection{Impact of neutrino masses\label{neutrinomass}}

In this subsection, we address the probably most well-motivated
extension of the basic cosmological model, massive neutrinos.  In
\cite{2007PhRvD..75b3522H}, the inclusion of a neutrino mass parameter
was shown to weaken parameter constraints on $n_{\rm S}$ and $r$
enough to affect conclusions about the $\lambda \phi^4$ model of
inflation in particular, mainly due to a parameter degeneracy with
$r$.  Interestingly we find that the more recent data effectively
break this degeneracy: the constraints for the vanilla+$r$+$m_\nu$
model in the ($n_{\rm S}$, $r$) plane are essentially identical to the
ones of the massless neutrino model, see top right panel of
Figure~\ref{fig:nsr}. We find an upper limit on the sum of neutrino
masses of $\sum m_\nu < 0.64\ \rm{eV}$ (at 95\% c.l.), consistent with
the results of a recent analysis with similar data
sets~\cite{2009arXiv0910.0008R}.  One might also wonder whether the
presence of massive neutrinos could help alleviate the problems of the
HZ model by virtue of the free-streaming-induced suppression of the
matter power spectrum at small scales.  This does not happen to be the
case here, however; neither the baryon density nor the spectral index
are significantly shifted with respect to the massless neutrino model.

\FIGURE{
\centering
  \includegraphics[height=.375\textwidth,angle=270]{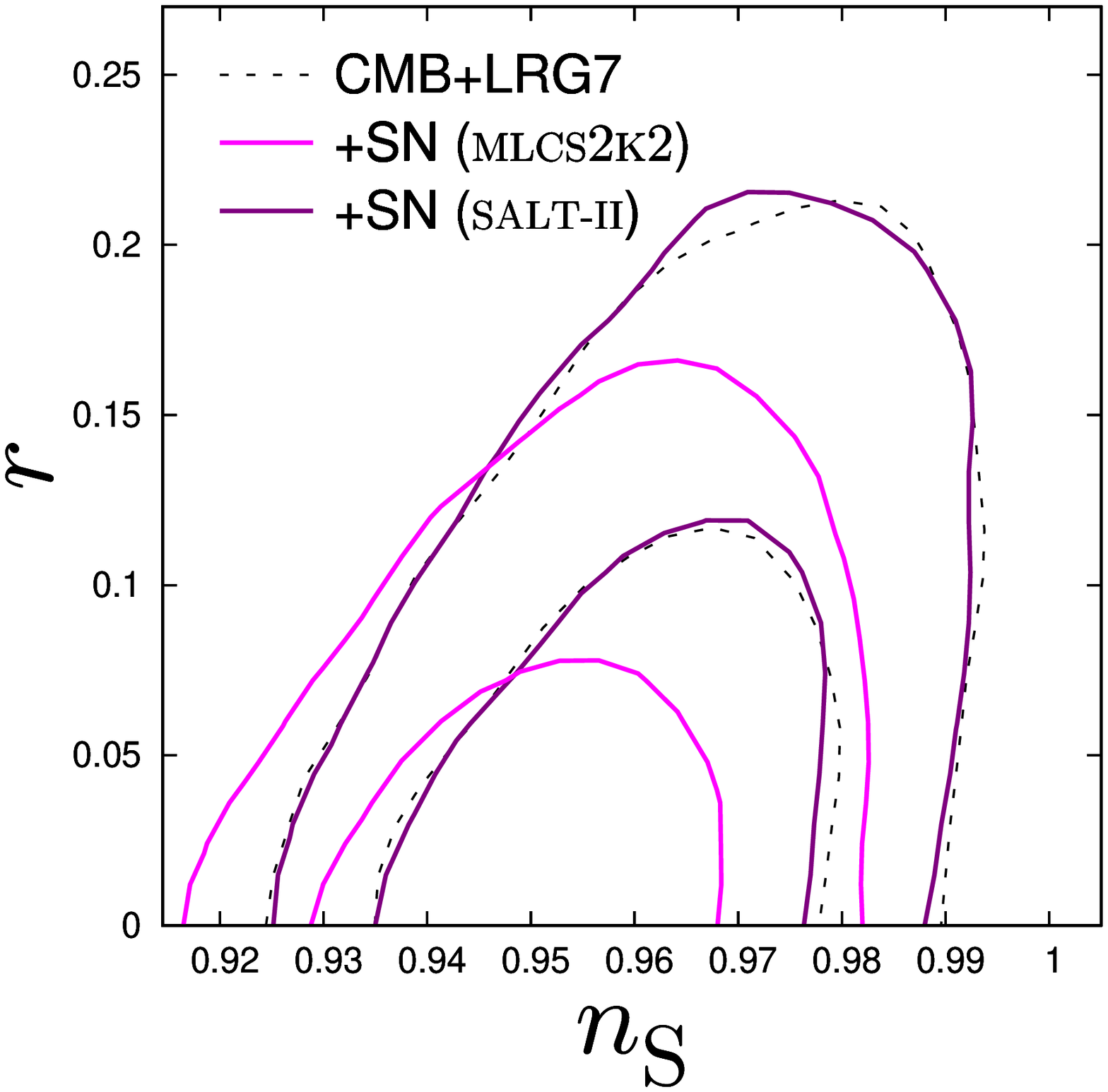}
  \includegraphics[height=.375\textwidth,angle=270]{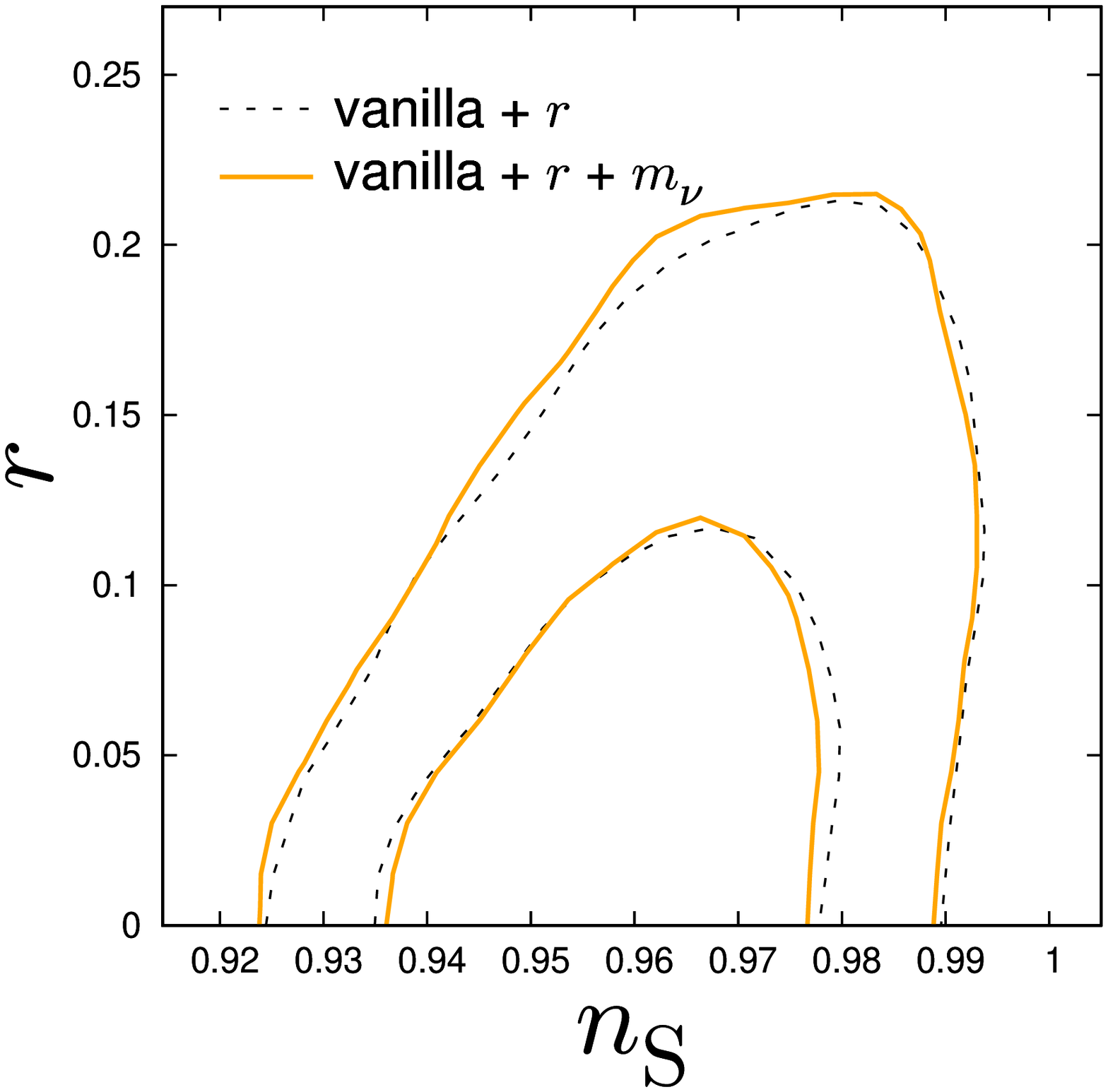}
  \includegraphics[height=.375\textwidth,angle=270]{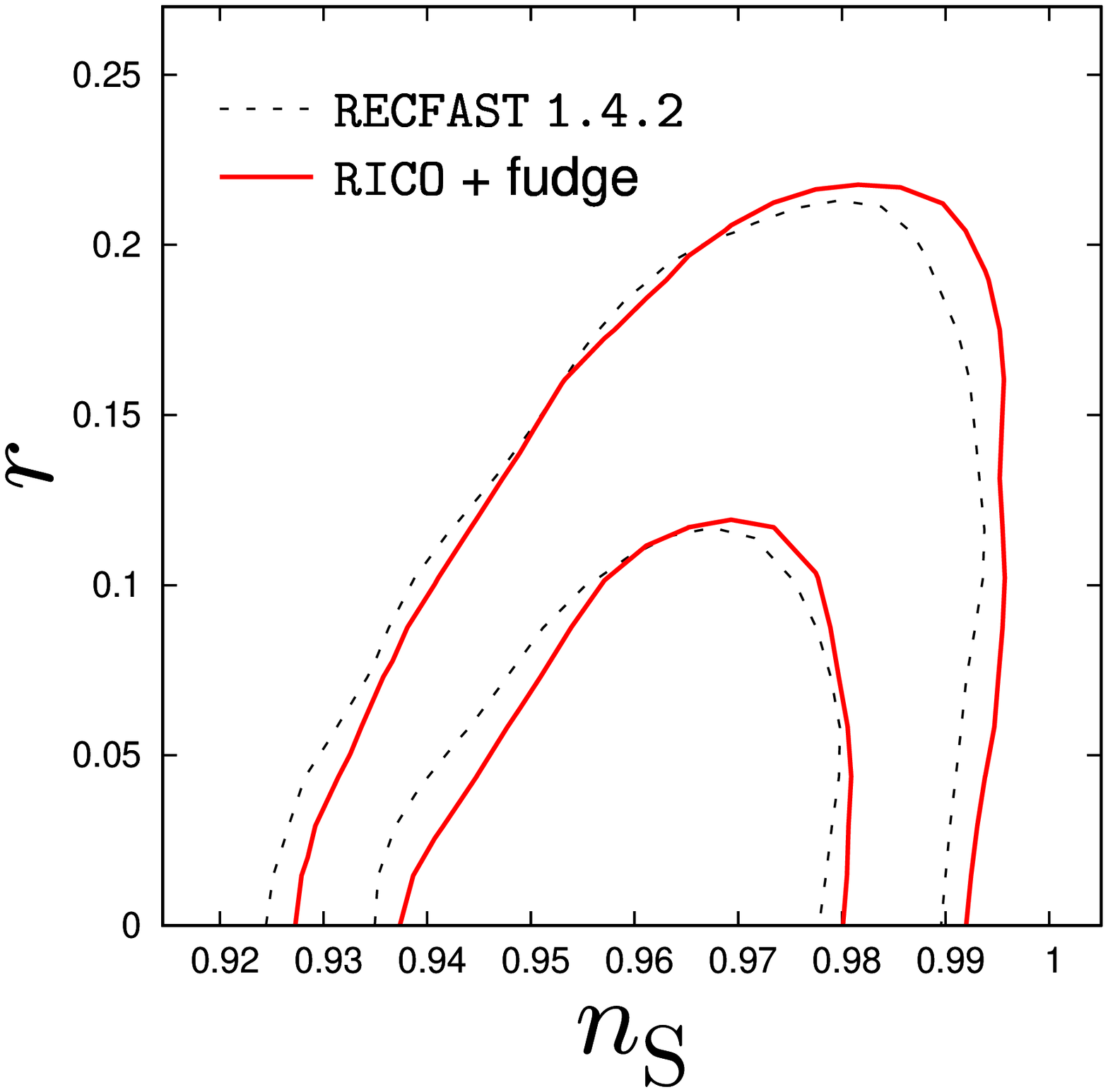}
  \includegraphics[height=.375\textwidth,angle=270]{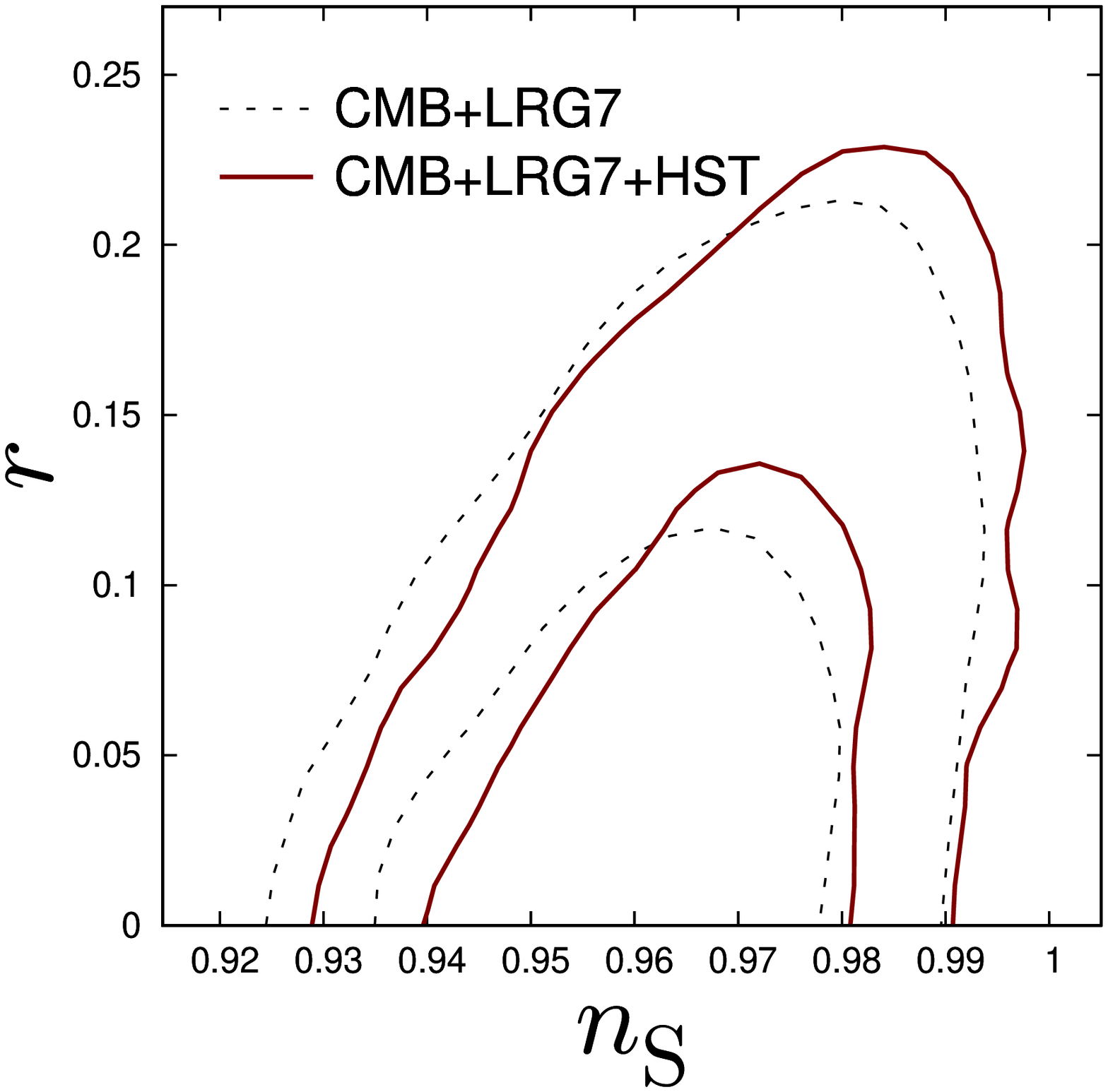}
  \caption{\label{fig:nsr} Marginalised joint 68\%- and 95\%-credible
    contours in the ($n_{\rm S}$, $r$) plane, illustrating the impact
    of various systematic effects.  The dotted black lines represent
    the results of our `default' setup in all three plots and are
    identical to the contours presented in Figure~\ref{fig:IVA}. 
    \emph{Top left:} Results for CMB+LRG7 data alone, and combined
    with luminosity distances from the supernova compilation of
    Ref.~\cite{Kessler:2009ys}, analysed with the {\sc mlcs2k2} (solid
    purple lines), and {\sc salt-ii} (solid pink lines) lightcurve
    fitters. 
    \emph{Top right:} Comparison of results for the massless neutrino
    and massive neutrino (orange lines) cases.   \emph{Lower left:}
    \texttt{CAMB}'s default recombination code \texttt{RECFAST 1.4.2}
    compared to recombination calculated with \texttt{RICO}, including
    a `fudge function' which accounts for corrections due to Raman
    scattering and Ly-$\alpha$ radiative transfer effects (red
    lines). \emph{Lower right:} Results for CMB+LRG7 compared to
    CMB+LRG7 combined with the $H_0$ measurement from
    \cite{2009ApJ...699..539R} (solid maroon lines).}
}

\subsection{Impact of recombination uncertainties}

One of the remaining major theoretical uncertainties in the
calculation of CMB angular power spectra is the physics of
recombination.  The standard approach relies on an effective
description of hydrogen recombination assuming a 3-level hydrogen
atom, as implemented in the widely-adopted \texttt{RECFAST} code
\cite{1999ApJ...523L...1S,2000ApJS..128..407S}, whose latest version
\texttt{1.4.2} \cite{2008MNRAS.386.1023W,2009MNRAS.397..445S}, is the
default model used in \texttt{CAMB}.  The authors of \texttt{RECFAST}
estimate that the computed ionisation fraction $x_{\rm e}(z)$ is
accurate at the percent level.  However, any error in $x_{\rm e}(z)$
will inevitably propagate to the $\mathcal{C}_\ell$s, affecting in
particular the small scales, and can potentially lead to biased
parameter estimates.

Recently, there has been a lot of progress in understanding the
physics of recombination (see for instance
Refs.~\cite{2009arXiv0910.4383R,2009arXiv0911.1359G}), and it seems
that the accuracy required for Planck-quality data has now been
reached.  The results of the latest state-of-the-art recombination
calculations can be implemented in a parameter estimation analysis
with the help of \texttt{RICO} \cite{2009ApJS..181..627F}, an
interpolation code for $x_{\rm e}(z)$.  In order to estimate the bias
from recombination errors on the inflation-related parameters, we
performed an analysis with \texttt{RICO} (including a `fudge function'
accounting for corrections due to Raman scattering and Ly-$\alpha$
radiative transfer effects, see Figure~1 of
Ref.~\cite{2009arXiv0910.4383R}) instead of \texttt{RECFAST}.

Our results are shown in the bottom panel of Figure~\ref{fig:nsr}; the
credible regions in the ($n_{\rm S}$, $r$) plane are shifted slightly
towards bluer tilts, though not at an alarming level.  We confirm the
conclusion of Ref.~\cite{2009arXiv0910.4383R} that for present data,
the current version of \texttt{RECFAST} is sufficiently accurate.

\section{Conservative constraints on the observable inflaton \mbox{potential}
\label{conservative}}

\subsection{Motivations}

The results of the previous section assume a negligible running of the
scalar and tensor tilts.  Indeed, these runnings are
  completely generic in inflation, but typically as small as
  $\alpha_{\rm S} \leq {\cal O}(10^{-4})$ for the simplest slow-roll
  models, as already stated in Section~\ref{simple}. 
In other words, we implicitly supposed in
  Section~\ref{simple} that the inflaton rolls down a smooth
potential well satisfying the slow-roll conditions and providing
enough inflationary $e$-folds for matching on to the post-inflationary
expansion era.  These assumptions maximise the beauty and simplicity
of the inflationary paradigm. However, nature sometimes prefers some
degree of complexity, and in principle inflation could take place
along a complicated potential, or in a multi-dimensional field space,
or in several stages associated to distinct mechanisms.  Even in that
case, inflation remains a key ingredient in the cosmological
evolution, necessary to solve the flatness and horizon problems and to
generate primordial perturbations.  If we take the point of view of
studying what the data really tells us on the inflaton potential (with
the only assumption that when cosmological scales leave the Hubble
scale, inflation is driven by a single scalar field with a canonical
kinetic term), the best we can do is to measure the primordial
spectrum or the underlying inflaton potential within the region probed
by cosmological observations, with no attempt to extrapolate beyond
that range.

It is sometimes argued that fitting the data with a large running in
the scalar spectrum (of the order of $\alpha_{\rm S} \sim {\cal
  O}(10^{-2})$ or more) is not consistent with the inflationary
paradigm, because -- as already argued in section \ref{motivnorunning}
-- such a large running signals that inflation ends soon after
observable scales leave the Hubble radius. It is worth clarifying this
argument. Of course, as long as the purpose is to constrain inflation
self-consistently, one should check that any large running model
providing a good fit to the data is indeed compatible with the
assumption that the expansion is accelerated as long as observable
scales leave the Hubble radius. If this is the case, such a model
cannot be eliminated without invoking some criteria of simplicity.
Indeed, the observable part of the potential can always be
extrapolated in such way to accommodate an arbitrary number of
$e$-folds. For models with a large tilt running in the scalar
spectrum, the {\it simplest} extrapolation schemes would suggest that
inflation ends soon after cosmological scales leave the Hubble scale,
but it is always possible to design the potential (i.e., to introduce
enough derivatives) in order to extend the duration of inflation as
desired. Alternatively, and as already mentioned in section
\ref{motivnorunning}, it is always possible to stick to the assumption
that the potential is smooth and inflation ends very quickly, provided
that later on, one or several extra inflationary stages (associated
with other inflatons) sum up to the desired 30 to 60 $e$-folds of
accelerated expansion. This situation can even be argued to be generic
in some particle physics frameworks see e.g. \cite{Adams:1997de}).

\subsection{Methods}

The shape of the inflationary potential within the
observable region can be constrained through various approaches that
make accurate inflationary predictions:
\begin{itemize}
\item The scalar primordial spectrum can be Taylor-expanded at various
  orders (including a running of the tilt, a running of the running,
  etc.).  Then, the free parameters of the model consist of the scalar
  spectrum amplitude, its various logarithmic derivatives at the pivot
  scale, and the tensor-to-scalar ratio. For consistency, the tensor
  spectrum shape should be fixed by the hierarchy of self-consistency
  conditions truncated at some order (see Appendix \ref{A1}). The data are
  used to provide constraints on these parameters ($A_{\rm S}$,
  $n_{\rm S}$, $\alpha_{\rm S}$, ..., $r$). A posteriori, these
  constraints can be converted into constraints on the scalar
  potential derivatives (or on combinations of these derivatives
  called the `potential slow-roll parameters') by making use of some
  analytical formula valid at a given order in slow-roll (with the
  caveat that slow-roll can be marginally satisfied in this context).
  This approach was chosen for instance by \cite{Komatsu:2008hk}.
\item One can choose a parametrisation of $V(\phi)$ in
  the observable range (if the `observable potential' is assumed to
  be smooth, a Taylor expansion is adequate). For each potential
  parameter set, the primordial spectra can be computed analytically
  or numerically and the model can be fitted to the data. In this
  case, the data provide direct constraints on the potential
  parameters, without relying on any slow-roll expansion in the case
  of a numerical computation. The result depends on the
  initial value of $\dot{\phi}$.  However, if one does not assume that
  inflation started just before the observable region is crossed,
  there is a unique choice for $\dot{\phi}$, corresponding to the
  inflationary attractor solution in phase space.  This approach can
  be easily followed by making use of the \texttt{CosmoMC} inflation
  module\footnote{\tt http://wwwlapp.in2p3.fr/\~{
    }valkenbu/inflationH/} released together with
  \cite{2007PhRvD..75l3519L}. This module performs a fully numerical,
  accurate and fast computation of the primordial spectra, based on
  the integration of background and perturbation equations during
  inflation (see Appendix \ref{A3}).
\item In order to relax the assumption on $\dot{\phi}$, one can
  perform a similar analysis targeting the function $H(\phi)$
  instead of $V(\phi)$. Since each $H(\phi)$ corresponds to a unique
  $V(\phi)$, this method is equally appropriate for constraining
  $V(\phi)$, and naturally incorporates models for which inflation
  starts just before entering the observable region. This approach was
  followed in \cite{Kinney:2002qn,Kinney:2006qm,Peiris:2006ug,Easther:2006tv,Peiris:2006sj,Peiris:2008be},
  with various analytical or semi-analytical schemes for
    the calculation of the primordial spectrum (based on an expansion
    of $H(\phi)$ into Hubble Slow-Roll (HSR) parameters). It was also
    followed in \cite{Lesgourgues:2007aa,Powell:2007gu,Hamann:2008pb}
    with a numerical calculation of the spectrum for Taylor-expanded
    functions $H(\phi)$. Although reconstructing $H(\phi)$ is
  slightly more general than $V(\phi)$, we will not follow this
  approach in the current paper because constraints obtained directly
  on the potential parameters (for instance on $V'/V$
    instead of $H'/H^2$) are slightly more convenient and suggestive
    for inflationary model-building.
\item One can try to reconstruct the function $H(N)$ where $N = \ln a$
  is the number of $e$-folds, using its expansion in Horizon Flow
  Functions (HFFs) evaluated at the time $N_*$ when the pivot scale
  crosses the Hubble radius during inflation. Combinations of the
  first three HFFs at the pivot scale can be related to the primordial
  spectra using the second-order slow-roll approximation
  \cite{Leach:2002ar}. The same approximation also provides relations
  with the derivatives of the inflaton potential (see Appendix
  \ref{A2}). This approach was chosen in many papers, for instance
  \cite{Leach:2003us, 2006JCAP...08..009M, Finelli:2006fi}.
\end{itemize}

In this work, we will mainly compare the first and second approaches,
although including also the fourth one for comparison.
Hence, we will first fit the data with primordial spectra described by
the set of parameters ($A_{\rm S}$, $n_{\rm S}$, $\alpha_{\rm S}$,
$r$), and use second-order slow-roll formulae to translate our results
either in terms of convenient combination of the potential parameters
$(V, V', V'', V''')$, or in terms of the HFF parameters. Next, we will
check whether these results agree with a direct reconstruction of the
HFF parameters (using the second-order slow-roll apprimation to relate
HFF parameters with the primordial spectra). Finally, we will relax
any slow-roll assumption and constrain directly the same quantities
using the numerical module of \cite{2007PhRvD..75l3519L}, assuming
that the observable potential $V(\phi)$ can be described by an 
order two or three Taylor-expansion.

\subsection{Results}

\subsubsection{Flat priors on the 
spectral parameters including running\label{prior1}}

We first run \texttt{CosmoMC} with flat priors on the spectral
parameters ($\ln A_{\rm S}$, $n_{\rm S}$, $\alpha_{\rm S}$, $r$).
For each parameter, the mean values and marginalised bounds on the
central 95\%-credible interval are given in Table~\ref{tab:r+alpha}
for three cases: with CMB data only, with our default dataset
CMB+LRG7, and for the same dataset when $r$ is kept fixed to zero
(low-scale inflation limit). As in previous papers on the subject 
(see for instance \cite{Finelli:2006fi,2007PhRvD..75l3519L}), a
large negative running is preferred, since the value
$\alpha_{\rm S}=0$ is always above the 95\%-credible interval. 
Adding the LRG7 data leads to a slightly smaller mean value
$\alpha_{\rm S}=-0.063$, but does not increase the level of
significance at which running spectra are preferred.  With the
CMB+LRG7 data set, introducing a running decreases the minimum
effective chi square by $\Delta (- 2 \ln({\cal L})) = 5.8$,
showing that a non-zero running is preferred by the
  data, but not with a high degree of
  significance\footnote{While this work was being
    completed, Ref.~\cite{2009arXiv0911.5191K} appeared, based on a
    slightly different data set.  By computing the `Bayesian evidence'
    ratio between the running and power-law models (with a top-hat
    prior on $\alpha_{\rm S}$ in the range [-0.1,0.1]), this analysis
    concludes that `running is not disfavored by the data nor required
    in modeling the data'.} (we recall that between the HZ and
  power-law model the effective chi square decreases by twice the same
  amount).  In Figure~\ref{fig:epsvsphys} (red
  curves), we show the marginalized likelihood contours for
  two-dimensional projections of the parameter space ($n_{\rm S}$,
  $\alpha_{\rm S}$, $r$).  As usual, the running is found to be
slightly correlated with the tensor-to-scalar ratio, and for large
values of $r$ the data are compatible with more running. However, in
the small-scale inflation limit (i.e., when we fix $r=0$), running is
preferred with roughly the same level of significance as in the $r
\neq 0$ model, since in that case we find
$\alpha_{\rm S}=-0.046^{+0.038}_{-0.039}$ at the 95\% level.

In the running model, the constraint on $r$ degrades by a factor 2
with respect to the power-law model. The allowed range for $n_{\rm 
  S}$ enlarges by a factor 1.5, but the mean value remains identical.
Indeed, we choose our pivot scale at $k_*=0.017$~Mpc$^{-1}$ in order to
remove most of the degeneracy between $n_{\rm  S}$ and
$\alpha_{\rm S}$, as can be checked in Figure~\ref{fig:epsvsphys}.

\FIGURE[h!]{
\centering
\includegraphics[width=.6\textwidth]{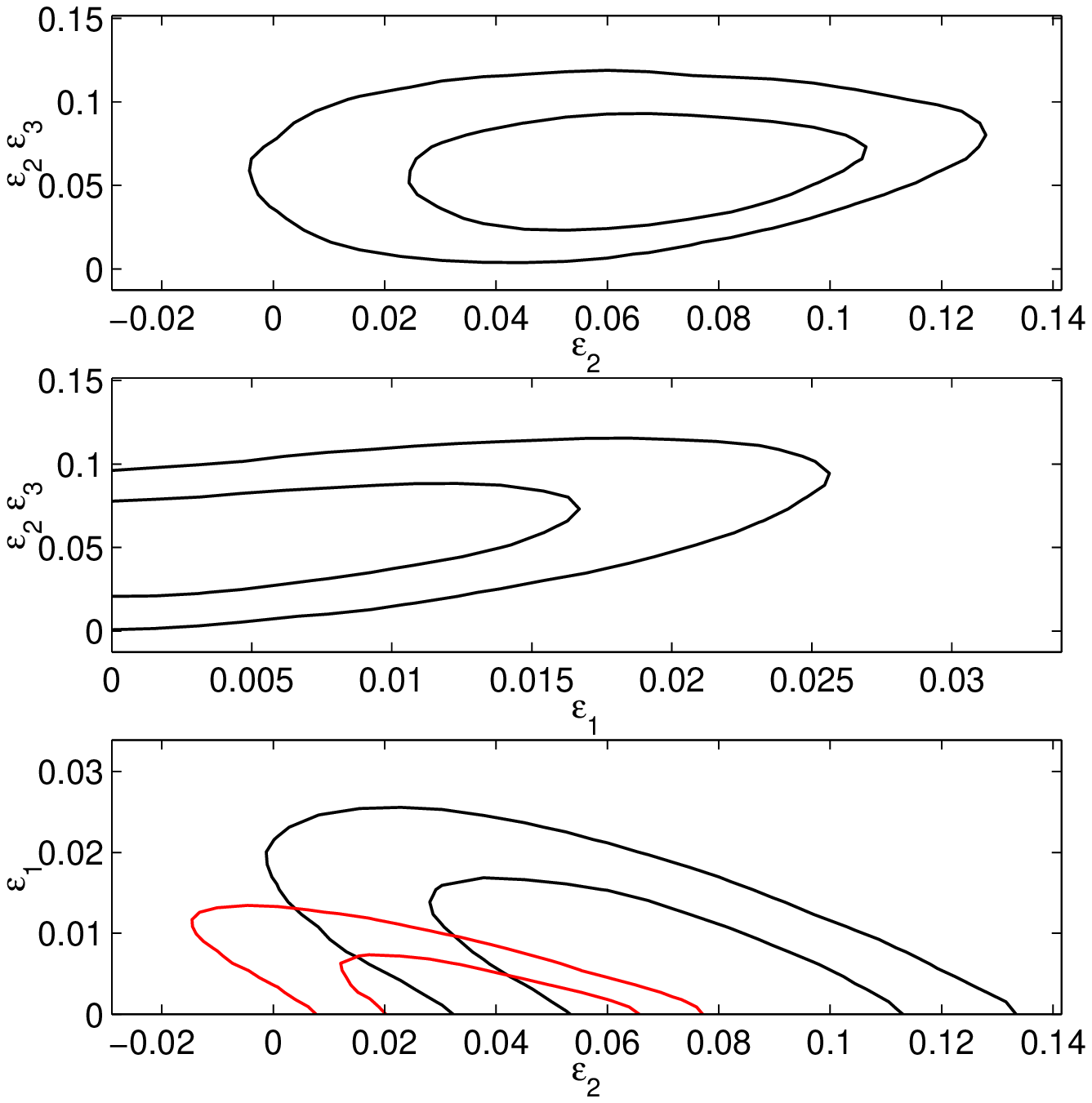}
\caption{Constraints from the CMB+LRG7 data set on combinations of the
  HFFs evaluated at the pivot scale ($\epsilon_1$, $\epsilon_2$,
  $\epsilon_2 \epsilon_3$) assuming: {\it (red)} $\epsilon_3=0$ and
  the first-order slow-roll approximation for the computation of the
  primordial spectra; {\it (black)} $\epsilon_3 \neq 0$ and the
  second-order slow-roll approximation.}
\label{fig:eps1eps2}
}
\subsubsection{Flat priors 
on combinations of the HFFs at the pivot scale\label{prior2}}

Next, for comparison, we run \texttt{CosmoMC} with flat priors on
combinations of the first three HFFs evaluated at the pivot scale,
namely $\epsilon_1$, $\epsilon_2$ and $\epsilon_2 \epsilon_3$ (see
Appendix \ref{A2} for definitions). We use the second-order slow-roll
approximation in order to relate these parameters to the spectral
parameters ($n_{\rm S}$, $\alpha_{\rm S}$, $r$). 
In Figure \ref{fig:eps1eps2}, we show our marginalized likelihood contours in
two-dimensional projections of the parameter space ($\epsilon_1$,
$\epsilon_2$, $\epsilon_2 \epsilon_3$). In Figure
\ref{fig:epsvsphys}, we compare our results with those of the previous
subsection, both expressed in the space of spectral
parameters. Clearly, the ensemble of models covered by the two
parametrisations is the same, and differences might be expected only
due to different priors. However, the contours of the two runs overlap
perfectly. This is an indication that in the range of models allowed
by the data, the expressions for ($n_{\rm S}$, $\alpha_{\rm S}$,
$r$) in terms of ($\epsilon_1$, $\epsilon_2$, $\epsilon_2 \epsilon_3$)
are dominated by linear terms, so that flat priors on one parameter
set are nearly equivalent to flat priors on the other one. The run of
this section does not bring new information on inflationary models but
provides a useful self-consistency check.

\FIGURE[h!]{
\centering
\includegraphics[width=.6\textwidth]{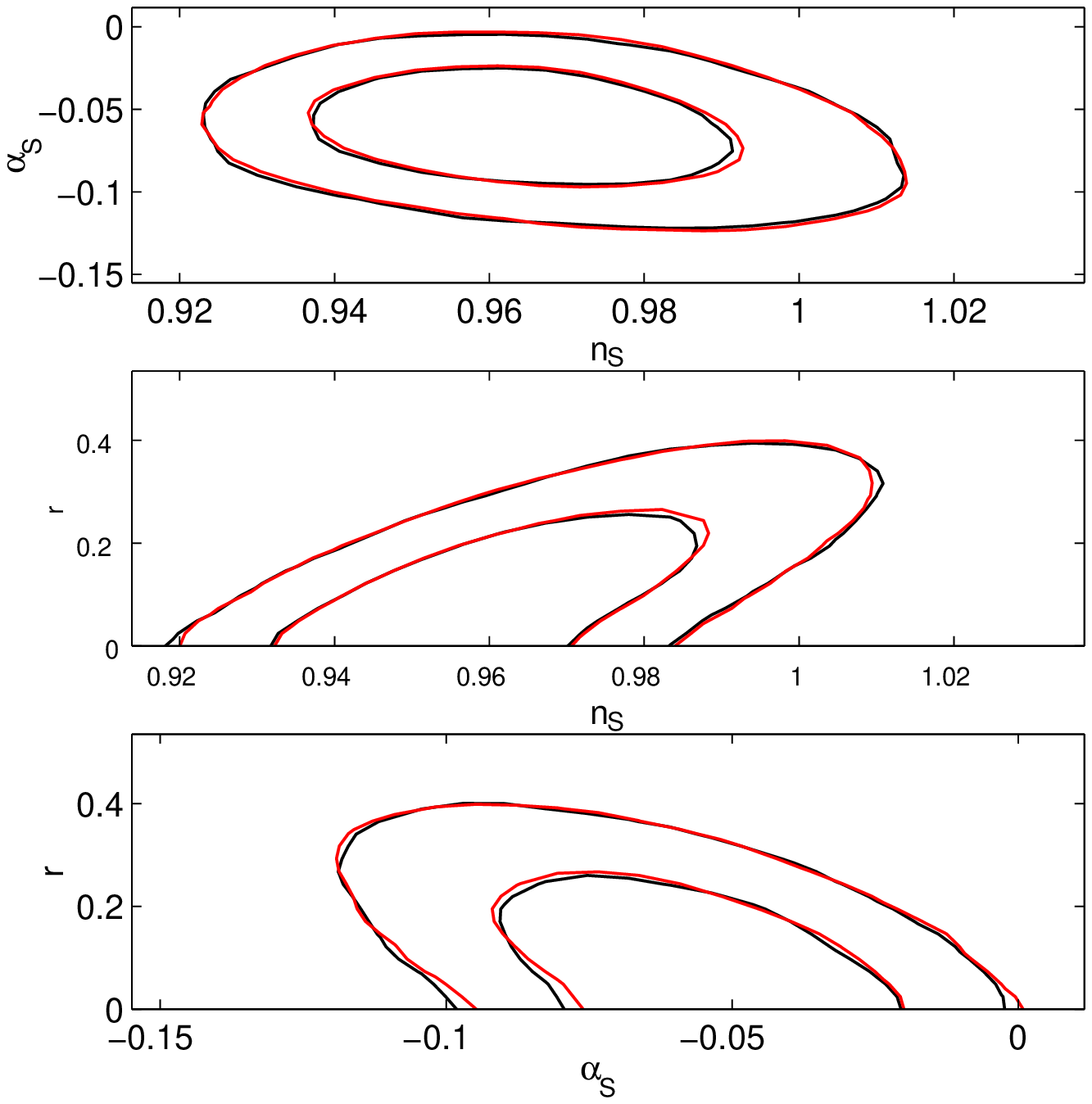}
\caption{Constraints from the CMB+LRG7 data set on combinations of the
  spectral parameters ($n_{\rm S} \,, \alpha_{\rm S} \,, r$). The
  black lines denote the constraints computed with the second-order
  slow-roll approximation, starting from flat priors on the HFF
  parameters at the pivot scale ($\epsilon_1$, $\epsilon_2$,
  $\epsilon_2 \epsilon_3$); the red lines denote the constraints
  obtained by using directly ($n_{\rm S} \,, r \,, \alpha_{\rm S}$),
  but enforcing the second-order consistency conditions for the
  tensor-to-scalar ratio and for the running of the tensor spectral
  index.}
\label{fig:epsvsphys}                                                 
}

Our constraints on the HFFs evaluated at the pivot scale are best
summarised by
\begin{equation}
\epsilon_1 < 0.021~,~~\epsilon_2 + 2.7 \epsilon_1 =
0.085^{+0.038}_{-0.039}~,~~\epsilon_2 \epsilon_3 =
0.061^{+0.046}_{-0.042}~~~~{\rm (95\%~c.l.).}
\end{equation}

When $\epsilon_3$ is fixed to zero, they reduce to
\begin{equation}
\epsilon_1 < 0.011~, \qquad
\epsilon_2 + 2.7 \epsilon_1 = 0.040\pm0.026~~~~{\rm ( 95\%~c.l.).}
\end{equation}

\subsubsection{Flat priors on combinations of the Taylor coefficients of the observable inflaton potential\label{prior3}}

The results of the last two subsections \ref{prior1}, \ref{prior2}
should be taken with a grain of salt for two reasons. First, when the
running is large, the second slow-roll approximation is not
necessarily accurate for all models allowed by the data, hence any
relation between the spectral parameters ($\ln A_{\rm S}$,
$n_{\rm S}$, $\alpha_{\rm S}$, $r$) and the underlying inflationary
potential $V(\phi)$ includes a theoretical uncertainty. Second, for
some spectra with a large running allowed by the data, there is no
guarantee that there exists a single underlying inflationary model
consistent with these spectra.  Since all allowed models have $r \ll
1$, it is clear that they are consistent with the condition
$\epsilon_1(k_*) \ll 1$ at the pivot scale, but on the edges of the
observable range, they might be incompatible with $\epsilon_1(k) \leq
1$, i.e., with inflationary expansion.

In the method consisting in fitting directly the potential $V(\phi)$
with a numerical computation of the primordial spectra over the
observable range, these caveats are avoided by construction.  Still,
one needs to make an assumption concerning the shape of the inflaton
potential in this range. In this subsection we assume that the
observable inflaton potential can approximated by a Taylor expansion
of order either two or three. Our free parameters are the potential
and its derivatives with respect to the inflaton field, evaluated when
the pivot scale $k_*$ crosses the Hubble radius during inflation:
$V_0=V$, $V_1\equiv dV/d\phi$, $V_2 \equiv d^2V/d\phi^2$ and $V_3
\equiv d^3V/d\phi^3$.  In order to avoid complicated parameter
degeneracies, we impose flat priors on the combinations $(V_1/V_0)^2$,
$V_2/V_0$ and $V_3 V_1 / V_0^2$: these parameters are related linearly
to the usual `potential slow-roll parameters' and, although no
slow-roll approximation is performed here, they remain not too far
from linear combinations of ($n_{\rm S}$, $\alpha_{\rm S}$, $r$)
\cite{2007PhRvD..75l3519L}.

\FIGURE[h!]{
\centering
\includegraphics[width=.6\textwidth]{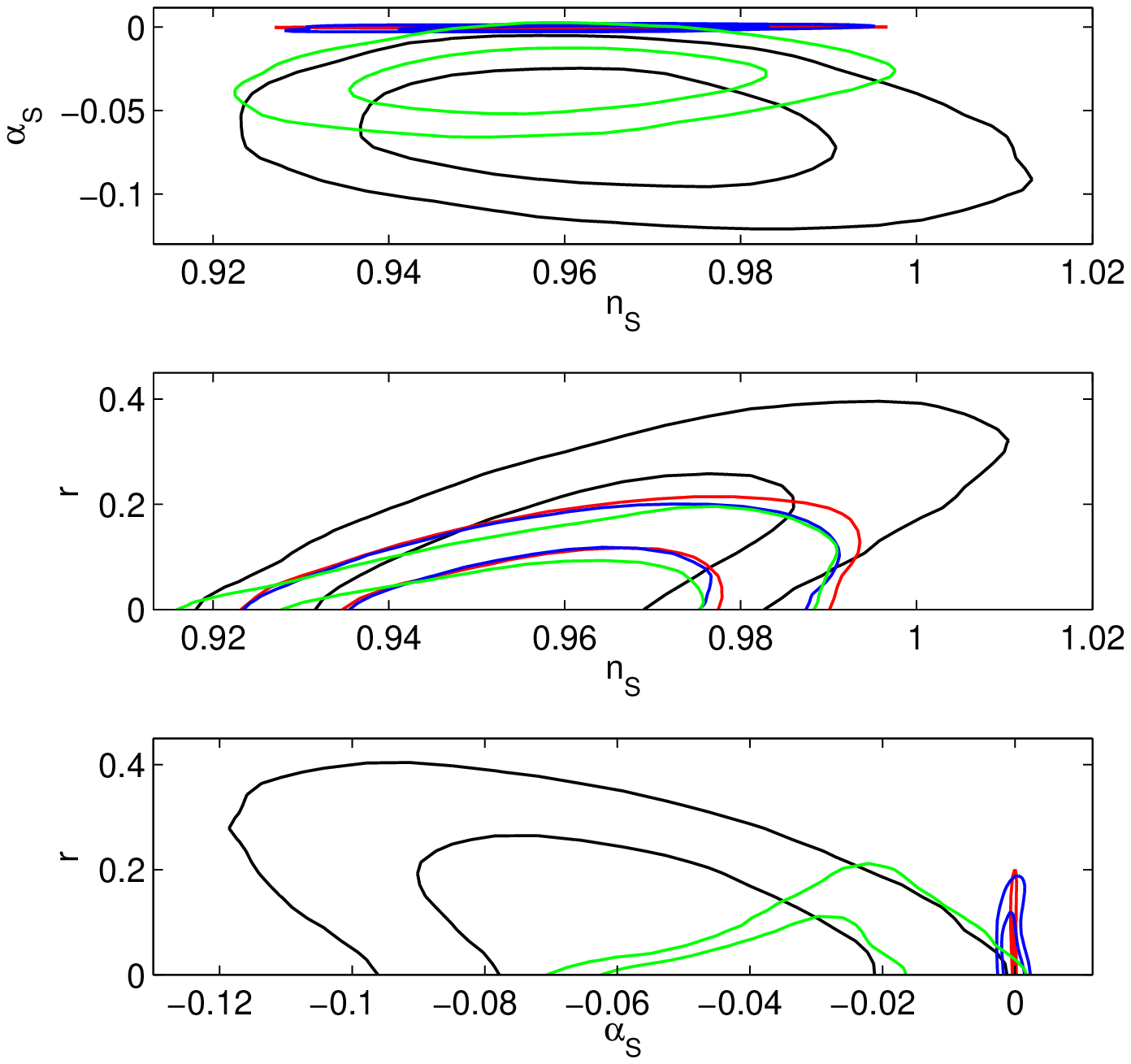}
\caption{Constraints from the CMB+LRG7 data set on combinations of the
spectral parameters ($n_{\rm S} \,, \alpha_{\rm S} \,, r$)
  under various assumptions: {\it (blue)} exact
  computation of the spectrum when the potential $V(\phi)$ is
  Taylor-expanded at order two in the observable range; {\it (green)}
  same with order three Taylor expansion; {\it (black)} spectrum
  computed with the second-order slow-roll approximation, starting
  from flat priors on the HFFs at the pivot scale
  ($\epsilon_1$, $\epsilon_2$, $\epsilon_2 \, \epsilon_3$); {\it
    (red)} same with $\epsilon_3=0$ prior (and first-order slow-roll
  approximation).}
\label{fig:phys123}
}

After running \texttt{CosmoMC}, for each model in our chains we can
compute ($n_{\rm S}$, $\alpha_{\rm S}$, $r$) defined at the pivot
scale directly from the numerical primordial spectra. Hence, the
results of this run can be readily compared in spectral parameter
space with those of the previous subsection based on HFF parameters
(like in \cite{2007PhRvD..75l3519L,2008JCAP...10..047A}). This
comparison is illustrated in Figure~\ref{fig:phys123}. As expected,
the run with $V_3=0$ leads to results very similar to those of the
power-law model with $\alpha_{\rm S}=0$. The reason is that a
quadratic potential cannot generate large running. This can be proved
using the second-order slow-roll approximation, which remains accurate
for all quadratic potentials compatible with the data. Following the
same logic, one could expect the run with $V_3 \neq 0$ to mimic the
results of the running model. This is far from being the case, as can
be seen in Figure~\ref{fig:phys123} when comparing the black and green
contours. The model with cubic terms in the potential cannot not reach
such large values of $|\alpha_{\rm S}|$ and $r$ as the running model.
This can be understood in the following way.  For models with large
running and tensors, the Horizon Flow Functions $\epsilon_1(N_*)$ and
$\epsilon_3(N_*)$ are not so small when the pivot scale crosses the
Hubble radius, and the flow of equations leads $\epsilon_1(N)$ to
reach order one even within the observable range. So, these models are
not consistent with any underlying potential described by a third
order polynomial within the observable range, even without assuming
any extrapolation scheme beyond this range. The limit between
consistent and inconsistent models depends crucially on the maximum
wavenumber $k_{\rm max}$ at which we require the primordial spectra to
converge after Hubble crossing (see Appendix \ref{A3} and the discussion in
\cite{Hamann:2008pb}), in our case $k_{\rm max}=5 {\rm Mpc}^{-1}$.
For this value and under the assumption that the observable inflaton
potential can be described by a third order polynomial, we find that
$r<0.15$ (at 95\% c.l.), i.e., roughly the same bound as in the
power-law model.  At the same time we get $\alpha_{\rm S}=-0.32 \pm
0.26$ (at 95\% c.l.): negative running is still favoured, but with
twice smaller values of $|\alpha_{\rm S}|$.

\FIGURE[h!]{
\centering
\includegraphics[width=.6\textwidth]{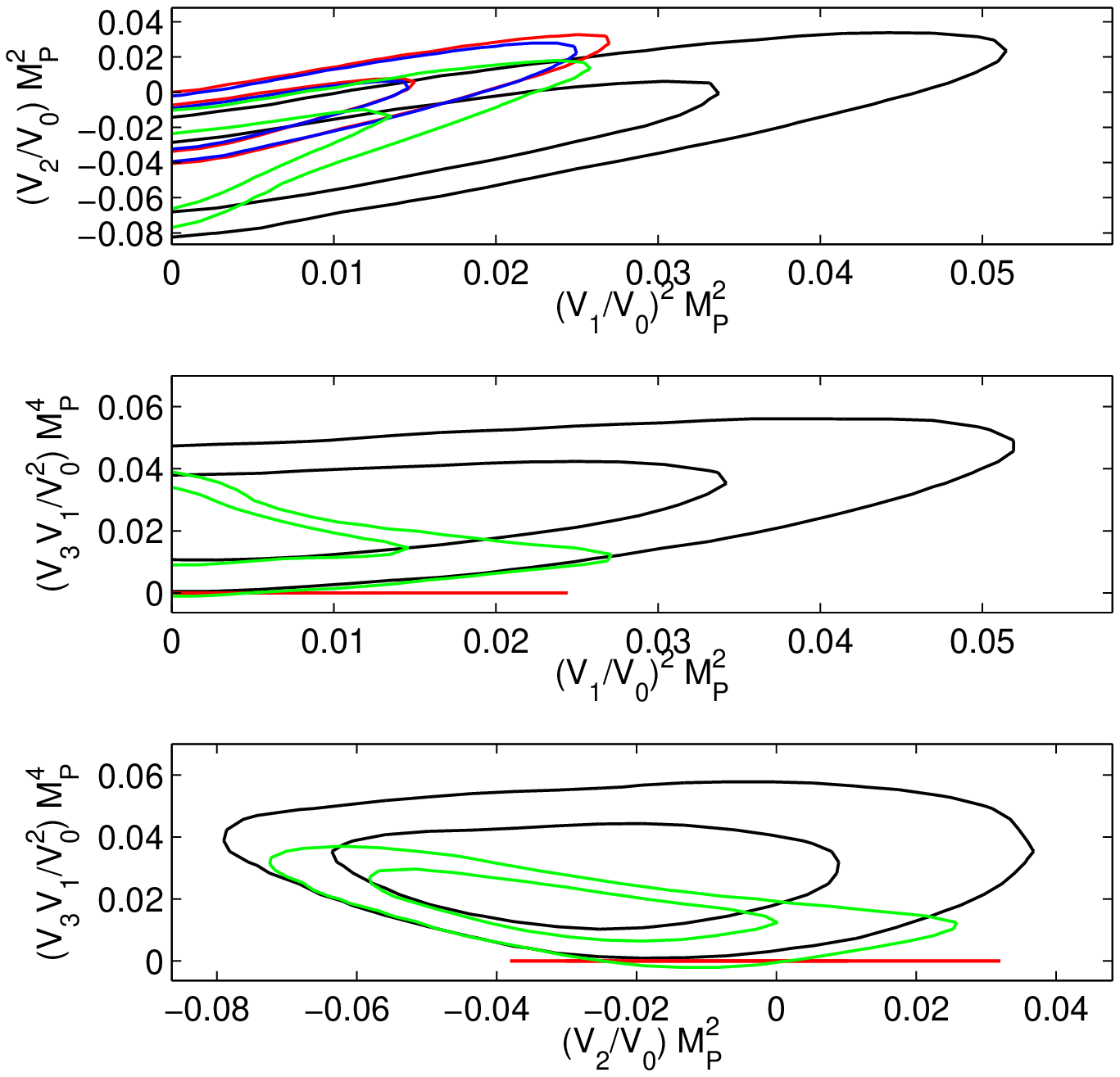}
\caption{Inflation constraints on combinations of the potential
  derivatives $V_1\equiv dV/d\phi$, $V_2 \equiv d^2V/d\phi^2$, $V_3
  \equiv d^3V/d\phi^3$ under various assumption: {\it (blue)} exact
  computation of the spectrum when the potential $V(\phi)$ is
  Taylor-expanded at order two in the observable range; {\it (green)}
  same with order three Taylor expansion; {\it (black)} spectrum
  computed with the second-order slow-roll approximation, starting
  from flat priors on the HFFs at the pivot scale
  ($\epsilon_1$, $\epsilon_2$, $\epsilon_2 \, \epsilon_3$), converted
  here into potential slow-roll parameters using exact formulae; {\it
    (red)} same with $\epsilon_3=0$ prior (and first-order slow-roll
  approximation). $M_{\rm P}$ stands for the reduced Planck mass.}
\label{fig:V123}
}

Note that the method of this subsection could be iterated at
higher-order in the Taylor expansion of the observable inflaton. The
results of \cite{2007PhRvD..75l3519L} based on older data suggest that
with a fourth-order expansion, the entire range of ($\alpha_{\rm S}$,
$r$) values allowed by the data in the running model could be
compatible with the $V(\phi)$ model. However, this is at the expense
of introducing a significant running of the running $\beta_{\rm S}$
(this parameter is governed by ${\rm d}^4V/{\rm d} \phi^4$ at the
pivot scale). In that case the allowed ranges for ($n_{\rm S}$,
$\alpha_{\rm S}$, $r$) would be even larger than found in section
\ref{prior1}.

The results of the runs based on potential parameters and HFF
parameters are also interesting to compare in the space of potential
parameters $(V_1/V_0)^2$, $V_2/V_0$ and $V_3 V_1 / V_0^2$. In order to
perform such a comparison, we re-mapped the HFF parameters into
potential slow-roll parameters using the formulae in Appendix \ref{A2}. 
The results are shown in Figure~\ref{fig:V123}. Again, the power-law model
and the second order expansion in $V(\phi)$ provide similar results,
as can be seen from the two-dimensional likelihood contours in
($(V_1/V_0)^2$, $V_2/V_0$) space. For these models the best
constrained combination is
\begin{equation}
\left[ \frac{V_2}{V_0} -1.9 \left(\frac{V_1}{V_0}\right)^2 \right] M_P^2 = 
-0.023\pm0.013~, ~~~~{\rm (95\%~c.l.)}
\end{equation}
while
\begin{equation}
\left(\frac{V_1}{V_0}\right)^2 M_P^2 < 0.020~. ~~~~{\rm (95\%~c.l.)}
\label{boundv1}
\end{equation}
When the third derivative of the potential is turned on, the combination
$V_3 V_1 /V_0^2$ is found to be in the range
\begin{equation}
\left(\frac{V_3 V_1}{V_0^2}\right) M_P^4 = 0.017^{+0.015}_{-0.014}~,
~~~~{\rm (95\%~c.l.)}
\end{equation}
while a reconstruction of the potential from spectral parameters (with
running) would suggest a larger range
\begin{equation}
\left(\frac{V_3 V_1}{V_0^2}\right) M_P^4 = 0.030^{+0.019}_{-0.018}~.~~~~{\rm (95\%~c.l.)}
\end{equation}
This discrepancy is equivalent to the one for $\alpha_{\rm S}$ discussed above.
For third-order polynomials $V(\phi)$, the upper bound on
$(V_1/V_0)^2$ is the same as in Eq.~(\ref{boundv1})
but the preferred combination of the first two parameter becomes
\begin{equation}
\left[ \frac{V_2}{V_0} -1.9 \left(\frac{V_1}{V_0}\right)^2 \right] M_P^2 = 
-0.041\pm0.021~. ~~~~{\rm (95\%~c.l.)}
\end{equation}

\TABLE{
\centering
\begin{tabular}{|l|ccc|}
\hline
& CMB only & CMB+LRG7 & CMB+LRG7 \\
& & & ($r=0$) \\
\hline
$\omega_{\rm  b}$ & $0.0221^{+0.0015}_{-0.0014}$ & $0.0219 \pm 0.0012$ & $0.0217 \pm 0.0012$ \\
$\omega_{\rm  c}$ & $0.116^{+0.017}_{-0.016}$    &   $0.120^{+0.009}_{-0.008}$ & $0.121^{+0.009}_{-0.008}$ \\
$h$ &  $0.677^{+0.076}_{-0.070} $ &  $0.694^{+0.038}_{-0.030}$ & $0.670^{+0.037}_{-0.035}$ \\
$\tau$ & $0.101 ^{+0.041}_{-0.037}$  &   $0.097^{+0.040}_{-0.035}$  &  $0.093^{+0.038}_{-0.034}$\\
$\log \left[ 10^{10} A_{\rm S}\right]$ & $3.17 \pm 0.12 $  & $3.18^{+0.09}_{-0.08}$ &  $3.18^{+0.09}_{-0.08}$\\
$n_{\rm  S}$ & $0.972^{+0.045}_{-0.037}$ & $0.964^{+0.039}_{-0.030}$ & $0.949^{+0.026}_{-0.025}$ \\
$r$ & $< 0.37$ & $< 0.33$ & -- \\
$\alpha_{\rm  S}$ & $-0.057 ^{+0.051}_{-0.054}$ & $-0.063^{+0.061}_{-0.049}$ &  $-0.046^{+0.038}_{-0.039}$\\

\hline
\end{tabular}
\caption{\label{tab:r+alpha}
  Mean parameter values and bounds of the central 95\%-credible intervals
  for the vanilla+$r$+$\alpha_{\rm S}$ and vanilla+$\alpha_{\rm S}$ parametrisations.}
}


\section{Conclusions\label{conclusions}}

We have provided a detailed update on the present observational status
of single-field inflation, paying particular attention to the
robustness of our results to selection of data sets and theoretical
uncertainties.  WMAP5 data, in combination with the measurements of
small scale CMB experiments and the halo power spectrum of luminous
red galaxies from the SDSS data release 7 provide increasingly tight
constraints on the physics of inflation.
After the completion of this work, the WMAP7 analysis of Komatsu et
al~\cite{2010arXiv1001.4538K} appeared and reached similar
conclusions. Still, some of the bounds presented here (based on WMAP5
data) are slightly tighter due to the inclusion of small-scale CMB
experiments and large scale structure data.

We find that evidence against the Harrison-Zel'dovich spectrum is
mounting, with a tilt of the spectrum now favoured at the level of
about four standard deviations.  Within the ($n_{\rm S}$,$r$)-model
which corresponds to most slow-roll inflation models, we infer
constraints of $n_{\rm S} = 0.962^{+0.028}_{-0.026}$ and $r<0.17$ (at
$95\%$ confidence level), which puts, e.g., chaotic $\lambda
\phi^4$-inflation under severe pressure, but leaves the quadratic
potential model inside the $95\%$~c.l. contours. Generally, the
tendency of the data to prefer a convex shape of the inflaton
potential has increased, though many concave models are still
viable. We have checked that these conclusions are robust with respect
to cosmological model extensions (such as the inclusion of a non-zero
neutrino mass) and to theoretical advances in modelling the physics of
recombination.
 
Tantalisingly, there remain mild indications that the data prefer a
negative running of the spectral index at roughly two standard
deviations -- indeed, a fair bit of allowed parameter space appears to
be in conflict with the predictions of models with negligible fourth
derivative of the potential in the observable range. We have also
presented a comparison among different methods in sampling the
inflationary parameters including running.

With a multitude of upcoming precise measurements of CMB polarisation
and smaller scale CMB temperature perturbations by experiments like
Planck, ACT, SPT, QUIET, SPIDER, PolarBear and EBEX, it is does not
seem too optimistic to expect that we will soon be able to resolve
these issues and take another big step towards reconstructing the
physics that governed the inflationary era of the Universe.

\section*{Acknowledgments}
We wish to thank Alberto Rubi\~no-Mart\'in, Alexei Starobinsky and
Wessel Valkenburg for very useful exchanges. We acknowledge the use of
the Legacy Archive for Microwave Background Data Analysis (LAMBDA).
Support for LAMBDA is provided by the NASA Office of Space Science.
Numerical computations were performed on the CINECA BCX cluster under
the INAF/CINECA agreement, the GRENDEL cluster at the Centre for
Scientific Computation Aarhus (CSC-AA), and the MUST cluster at LAPP
(CNRS \& Universit\'e de Savoie). FF thanks the CERN Theory division
for support. JH was supported by a Feodor Lynen-fellowship of the
Alexander von Humboldt foundation. JL and JH acknowledge support from
the EU 6th Framework Marie Curie Research and Training network
`UniverseNet' (MRTN-CT-2006-035863). This research was partially
supported by ASI contract I/016/07/0 `COFIS'.

\appendix
\section{Inflation predictions}
\addtocounter{subsection}{0}

We compare inflationary predictions with observations, by adopting a
parametrisation of the primordial power spectra (PS henceforth) of
curvature and tensor perturbations as:
\begin{equation}
\label{plex}
\ln{\mathcal{P}_X (k) \over \mathcal{P}_{X} (k_*)} = b_{0 \, X} + b_{1 \, X} \ln \left(k\over k_*\right)
+ \frac{b_{2 \, X}}{2} \ln^2\left(k\over k_*\right) 
\end{equation}
where $X={\rm S \,, T}$ stands for scalar and tensor, respectively,
$k_*$ is the pivot scale, $b_{1 \, S} = n_{\rm S} -1$, $b_{1 \,
  {\rm T}} = n_{\rm T}$, $b_{2 \, {\rm S}} = \alpha_{\rm S}$,
$b_{2 \, {\rm T}} = \alpha_{\rm T}$.

\subsection{Sampling the spectral parameters\label{A1}}

In Section~\ref{simple} we have used the parametrisation of
Eq.~(\ref{plex}), sampling directly from the logarithm of $A_{\rm S}$ 
($=\mathcal{P}_{\rm S}(k_*) e^{b_{0 \, {\rm S}}}$), $n_{\rm S}$ and $r$ 
($=\mathcal{P}_{\rm T}(k_*) e^{b_{0 \, {\rm T}}-b_{0 \, {\rm S}}}/\mathcal{P}_{\rm S}(k_*)$), 
with \mbox{$b_{2 \, X}=0$}.  The tensor spectral index has
been fixed through the (first-order) consistency condition:
\begin{equation}
n_{\rm T} = -\frac{r}{8} \,.
\end{equation}

In Section~\ref{conservative} we have sampled $A_{\rm S}$,
$n_{\rm S}$, $r$, $\alpha_{\rm S}$. The tensor spectral index has
been fixed through the (second-order) consistency condition:
\begin{equation}
n_{\rm T} = -\frac{r}{8} \left( 2 - \frac{r}{8} - n_{\rm S} \right) \,.
\end{equation}
The running of the tensor spectral index has been fixed to:
\begin{equation}
\alpha_{\rm T} = \frac{r}{8} \left( \frac{r}{8} + n_{\rm S}-1 \right) \,.
\end{equation}

\subsection{Sampling the Horizon Flow Functions (HFFs)\label{A2}}

In Section~\ref{conservative} we have derived $\mathcal{P}_X(k_*) \,,
b_{i \, X}$ from the Hubble parameter $H$ and the {\em horizon flow
  functions} $\epsilon_i$ (HFF henceforth) evaluated at the pivot
scale $k_*$.  The HFFs are defined as $\epsilon_1 = - \dot H/H^2$ and
$\epsilon_{i + 1} \equiv \dot \epsilon_i/(H \epsilon_i) = ({\rm d}
\epsilon_i/{\rm d} \, N)/\epsilon_i$ with $i \ge 1$ and $N$ the number
of $e$-folds (${\rm d} N = H {\rm d}t$) \cite{Schwarz:2001vv}.  The
analytic slow-roll approximated power spectra has been obtained first
through the Green's function method (GFM henceforth) in
Refs.~\cite{Gong:2001he,Leach:2002ar}. The coefficients for the scalar
spectrum are:
\begin{eqnarray}
\label{eqn:bs0}
b_{{\rm S}0} &=&
- 2\left(C + 1\right)\epsilon_1 - C \epsilon_2
+ \left(- 2C + {\textstyle\frac{\pi^2}{2}} - 7\right)              
 \epsilon_1^2 + \left({\textstyle\frac{\pi^2}{8}} - 1\right) 
\epsilon_2^2 \nonumber\\& &
+ \left(- C^2 - 3C + {\textstyle\frac{7\pi^2}{12}} - 7 \right)
\epsilon_1\epsilon_2
+ \left(-{\textstyle\frac 12}C^2 +
{\textstyle\frac{\pi^2}{24}} \right)
\frac{{\rm d} \epsilon_2}{{\rm d} N} \, \\
\label{eqn:bs1}
b_{{\rm S}1} &\equiv&  n_{\rm S} - 1 = - 2 \epsilon_1 - \epsilon_2 - 2
\epsilon_1^2 -\left(2\,C+3\right)\,\epsilon_1\,\epsilon_2 - 
C\, \frac{{\rm d} \epsilon_2}{{\rm d} N} \, \\
b_{{\rm S}2} &\equiv& \alpha_{\rm S} = - 2 \epsilon_1 \epsilon_2 -
\frac{{\rm d} \epsilon_2}{{\rm d} N}
\label{eqn:bs2},
\end{eqnarray}
and for tensors are:
\begin{eqnarray}
b_{{\rm T}0} &=&
 - 2\left(C + 1\right)\epsilon_1
 + \left(- 2C + {\textstyle\frac{\pi^2}{2}} - 7\right)
 \epsilon_1^2
+ \left(-C^2 - 2C + {\textstyle\frac{\pi^2}{12}} - 2 \right)
 \epsilon_1\epsilon_2 ,
\label{eqn:bt0}
\\
b_{{\rm T}1} &\equiv&  n_{\rm T} = - 2\epsilon_1 - 2\epsilon_1^2
-2\,\left(C+1\right)\,\epsilon_1\,\epsilon_2 , \\                  
\label{eqn:bt1}
b_{{\rm T}2} &\equiv&  \alpha_{\rm T} = - 2\epsilon_1\epsilon_2,
\label{eqn:bt2}
\end{eqnarray}
where $C \equiv \ln 2+\gamma_{\rm E}-2\approx-0.7296$ ($\gamma_{\rm
  E}$ is the Euler-Mascheroni constant) and ${\rm d} \epsilon_2 / {\rm d} N =
\epsilon_2 \epsilon_3$. A similar structure has been obtained through
the method of Comparison Equation \cite{Casadio:2006wb}.

For a Klein-Gordon scalar field $\phi$, the potential $V$ and its
derivatives are related to $H$ and $\epsilon_i$ as: 
\begin{eqnarray}
V &=& 3 \, M_{\rm P}^2 \, H^2 \left(1 -
\frac{\epsilon_1}{3}\right) \,, \nonumber \\
\frac{V_\phi^2 M_{\rm P}^2}{V^2} &=& 2 \epsilon_1 
\frac{\left(1 - \frac{\epsilon_1}{3} +
\frac{\epsilon_2}{6}\right)^2}{\left(1 -
\frac{\epsilon_1}{3}\right)^2} \,, 
\nonumber \\
\frac{V_{\phi \phi} M_{\rm P}^2}{V} &=& \frac{ 2 \epsilon_1 - \frac{\epsilon_2}{2}
- \frac{2\epsilon_1^2}{3} + \frac{5\epsilon_1 \epsilon_2}{6}
-\frac{\epsilon_2^2}{12} - \frac{\epsilon_2 \epsilon_3}{6}}{1 -
\frac{\epsilon_1}{3}} \,, \nonumber \\
\frac{V_{\phi \phi \phi} V_\phi M_{\rm P}^4}{V^2} &=&
\frac{1 - \frac{\epsilon_1}{3} + 
\frac{\epsilon_2}{6}}{\left(1-\frac{\epsilon_1}{3}\right)^2} 
\left( 4 \epsilon_1^2 -3 \epsilon_1 \epsilon_2 
+ \frac{\epsilon_2 \epsilon_3}{2} 
+ \epsilon_1^2 \epsilon_2 - \epsilon_1 \epsilon_2^2 \right. \nonumber \\
&& - \left.\frac{4}{3} \epsilon_1^3  - \frac{7}{6} \epsilon_1 \epsilon_2 \epsilon_3
+ \frac{\epsilon_2^2 \epsilon_3}{6}
+ \frac{\epsilon_2 \epsilon_3^2}{6}
+ \frac{\epsilon_2 \epsilon_3 \epsilon_4}{6}
\right)  \,.
\label{potential_eps}
\end{eqnarray}
The use of the HFF parametrisation allows to reconstruct the
derivatives of the potential without any additional approximation. 

\subsection{Sampling the potential parameters\label{A3}}

In Section~\ref{conservative} we used the \texttt{CosmoMC}
inflationary module\footnote{\tt http://wwwlapp.in2p3.fr/\~{
  }valkenbu/inflationH/} released together with
\cite{2007PhRvD..75l3519L}. The basic principles of this module are the
following.

In \texttt{CosmoMC}, the pivot scale is fixed once and for all, but
different parameters defining the function $V(\phi-\phi_*)$ are passed
to \texttt{CAMB} (here $\phi_*$ is the value of the inflaton field
when the pivot scale crosses the Hubble radius; it does not need to be
formulated explicitly). For each $V(\phi-\phi_*)$, the module
computes the spectra $\mathcal{P}_{\rm S}(k)$, $\mathcal{P}_{\rm T}
(k)$ within the range $[k_{\rm min}, k_{\rm max}]=[5\times
10^{-6},5]~$Mpc$^{-1}$ needed by \texttt{CAMB}, imposing that $aH=k_*$
when $\phi=\phi_*$. So, the code first finds the inflationary
attractor solution around \mbox{$\phi=\phi_*$}, computes $H_*$ and normalizes
the scale factor so that $a_*=k_*/H_*$. Then, each mode is integrated
numerically for $k/aH$ varying between two adjustable ratios: here,
$50$ and $1/50$. The evolution of each scalar/tensor mode is given by
\begin{equation}
    \frac{{\rm d}^2 \xi_{\rm S,T}}{{\rm d}\eta^2} + \left[ k^2 - \frac{1}{z_{\rm S,T}}
    \frac{{\rm d}^2 z_{\rm S,T}}{{\rm d}\eta^2}\right] \xi_{\rm S,T}=0
\end{equation}
  with $\eta=\int {\rm d}t/a(t)$ and $z_{\rm S}= a\dot\phi/H$ for scalars, 
  $z_{\rm T}= a$ for tensors.
  The code integrates this equation starting from the initial condition
  $\xi_{\rm S,T} = e^{-ik \eta}/\sqrt{2k}$ when $k/aH=50$, and computes
\begin{equation}
  \mathcal{P}_{\rm S} = \lim_{k \ll aH}
  \frac{k^3}{2 \pi^2} \frac{|\xi_{\rm S}|^2}{z_{\rm S}^2}~,
  \qquad
  \mathcal{P}_{\rm T} = \lim_{k \ll aH}
  \frac{32 k^3}{\pi m_{\rm P}^2} \frac{|\xi_{\rm T}|^2}{z_{\rm T}^2}~.
\end{equation}
So, the earliest (latest) time considered in the code is that when
$k_{\rm min}/aH=50$ ($k_{\rm max}/aH=1/50$), which in the attractor
solution uniquely determines extreme values of $(\phi-\phi_*)$
according to some potential. In the module this is translated to
demanding that $aH$ grows according to the aforementioned ratios: by
$50 k_*/ k_{\rm min}$ before $\phi=\phi_*$, and by $50 k_{\rm
  max}/k_*$ afterwards.  Hence, one of the preliminary tasks of the
module is to find the earliest time.  If by then, a unique attractor
solution for the background field cannot be found within a given
accuracy (10\% for $\dot{\phi}_{\rm ini}$), the model is rejected. So,
the module implicitly assumes that inflation starts at least a few
$e$-folds before the present Hubble scale exits the horizon (this
assumption is relaxed in the second version of the module based on
$H(\phi)$ reconstruction \cite{Lesgourgues:2007aa}).  In addition, the
module imposes a positive, monotonic potential and an accelerating
scale factor during the period of interest.  As a result of the chosen
method, the potential is slightly extrapolated beyond the observable
window, in order to reach the mentioned conditions for the beginning
and ending of the numerical integration. The range of extrapolation is
still very small in comparison with an extrapolation over the full
duration of inflation after the observable modes have exited the
Hubble radius.  Note that in this approach one does not need to make
any assumption about reheating and the duration of the radiation era.


\bibliography{fhll_jcap}

\end{document}